\documentclass[
    aps,
superscriptaddress,
    pr-applied, 
    twocolumn,
    letterpaper,
    10pt,
    floatfix
]{revtex4-1}

\usepackage{enumerate,graphicx,amsmath,amssymb}
\usepackage{epstopdf}
\usepackage{siunitx}
\usepackage{xcolor}
\usepackage{wrapfig}
\usepackage{float}
\usepackage{soul}

\def\beq{\begin{equation}}
\def\eeq{\end{equation}}
\newcommand{\delete}[1]{}

\newcommand{\be}{\begin{equation}}
\newcommand{\ee}{\end{equation}}

\newcommand{\fixme}{\textcolor{black}}

\begin{document}

\title{Nanomechanical testing of silica nanospheres for levitated optomechanics experiments}

\author{Cayla R. Harvey}
\affiliation{Department of Chemical and Materials Engineering, University of Nevada, Reno, NV 89557.}
\author{Evan Weisman}
\affiliation{Center for Fundamental Physics, Department of Physics and Astronomy, Northwestern University, Evanston, IL 60208.}
\author{Chethn Galla}
\affiliation{Center for Fundamental Physics, Department of Physics and Astronomy, Northwestern University, Evanston, IL 60208.}
\author{Ryan Danenberg}
\affiliation{Department of Physics, University of Nevada, Reno, NV 89557.}
\author{Qiyuan Hu}
\affiliation{Department of Physics, Williams College, Williamstown, MA 01267.}
\author{Swati Singh}
\affiliation{Department of Electrical and Computer Engineering, University of Delaware, Newark, DE 19716.}
\author{Andrew A. Geraci}
\affiliation{Center for Fundamental Physics, Department of Physics and Astronomy, Northwestern University, Evanston, IL 60208.}
\author{Siddhartha Pathak}
\affiliation{Department of Materials Science and Engineering, Iowa State University, Ames, IA 50011.}
\begin{abstract}
Optically-levitated dielectric particles can serve as ultra-sensitive detectors of feeble forces and torques, as tools for use in quantum information science, and as a testbed for quantum coherence in macroscopic systems. Knowledge of the structural and optical properties of the particles is important for calibrating the sensitivity of such experiments. Here we report the results of nanomechanical testing of silica nanospheres and investigate an annealing approach which can produce closer to bulk-like behavior in the samples in terms of their elastic moduli. These results, combined with our experimental investigations of optical trap lifetimes in high vacuum at high trapping-laser intensity for both annealed and as-grown nanospheres, were used to provide a theoretical analysis of the effects of porosity and non-sphericity in the samples, identifying possible mechanisms of trapping instabilities for nanospheres with non-bulk-silica-like properties.

\end{abstract}
\maketitle

\section{Introduction} 
\fixme{Optically-levitated} dielectric objects in ultra-high vacuum exhibit an excellent decoupling from their environment, making them highly promising systems for precision sensing and quantum information science. In particular, the center of mass modes of optically-trapped silica nanospheres have exhibited high mechanical quality factors in excess of $10^7$ \cite{novotny2012} and zeptonewton ($10^{-21}$ N) force sensing capabilities \cite{ranjit2016}. Such devices make promising candidates for sensors of extremely feeble forces \cite{geraci2010}, accelerations \cite{andyhart2015, Moore2017,novotnydrop}, torques \cite{Li2016}, and rotations \cite{Li2018,Novotny2018,Moore2018}, testing foundational aspects of quantum mechanics \cite{oriol2011}, observing quantum behavior in the vibrational of modes of mechanical systems \cite{chang2009,coherentscattering, aspelmeyercavity}, and tools for quantum information science perhaps especially when coupled to other quantum systems \cite{sympcool}.  

\fixme{Optically-trapped} dielectric nanoparticles can be used for a variety of fundamental physics experiments, including searches for micron-scale deviations from Newtonian gravity \cite{geraci2010}, tests for millicharged particles \cite{millicharge}, and searches for {\color{black} high frequency gravitational waves \cite{GWprl,LSDpaper}, dark matter \cite{LSDpaper,DMcarney, Carney2021} an dark energy \cite{Rider2016, Betz2022}. }
A number of recent theories including string theory, supersymmetry, and theories of ``large'' extra dimensions suggest the gravitational inverse square law may acquire a new form at sub-millimeter distances \cite{GiudiceDimopoulos,add}. As the gravitational force between massive objects becomes weak very rapidly as their size and separation distance decreases, ultra-precise measurements are a necessity at sub-millimeter length scales, and previous experiments have employed sensitive torsion balances \cite{Kapner2007}, cryogenic microcantilevers \cite{Geraci2008}, and torsional oscillators \cite{Chen2016}. Optically trapped spheres can {\color{black} also} function as a test mass 
in the search for Non-Newtonian gravity-like forces and Casimir forces.

For such precision sensing experiments it is often desirable to know the density and mass of the levitated particle, for example to properly determine the gravitational forces exerted on these sensors by nearby objects or by gravitational waves. Silica is often a material of choice for the levitated particle due to low optical absorption and laser-induced heating \cite{silicaglass}.  Previous data has suggested that the silica nanospheres may not be of solid density and thus could potentially contain voids \cite{ranjit2016,Moore2017}.  Such voids are deleterious for the proper estimation of gravitational forces and result in unknown effects on the scattering of light in the laser trap. 

In addition, several groups studying levitated opto-mechanical systems have observed a variety of trapping instabilities which occur at high vacuum or while pumping to high vacuum conditions, and some of these may be related to the material properties of the silica glass \cite{ranjit2016}. 
In Ref. \cite{ranjit2016}, the trapping lifetime in high vacuum for $300$ nm silica particles levitated by a 1064 nm laser was studied. At higher intensity there was an observed exponential reduction of lifetime with increasing laser power.  The estimated time scale to reach thermal equilibrium in these experiments is less than one second, despite lifetimes ranging from minutes to a few hours.  The exact loss mechanism is uncertain, but it has been suggested that a particle may undergo annealing or a glass-crystalline transition after remaining at an elevated temperature for a significant time. This could be responsible for \fixme{ejection of the particle from the optical trap} if the new phase has higher absorption or if the bead experiences a kick due to a sudden change in density, size, \fixme{mass}, or refractive index. Annealing is reported for certain forms of silica at temperatures as low as $500$ K over $30$-minute time scales \cite{bruckner}.

In Ref. \cite{Moore2017}, larger ($\sim 10$ micron) optically trapped microspheres had their radius become significantly reduced (e.g. by as much as $\sim 30 \%$) as the
pressure is quickly reduced from $\sim 0.1$ mbar to $10^{-5}$ mbar using a turbopump. Here the effect was attributed to heating of the particles in high vacuum conditions where gas collisions can no longer efficiently remove the heat produced by the laser absorption in the particle. 

Measurements of the elastic modulus of the nanoparticles and microspheres both before and after annealing could be beneficial for explaining these phenomena. However, the small (nanometer) length scale of the silica nanospheres makes it challenging to measure their elastic moduli, making \textit{in-situ} nanomechanical testing the tool of choice.  

In this paper, we report the characterization of the elastic modulus of the silica nanoparticles used in optical levitation experiments for a variety of samples and annealing preparation conditions.  We also report the results of optical trapping stability tests with annealed and as-grown nanospheres, and include a theoretical analysis of the expected perturbation on the optical 
{\color{black} trapping forces} which occurs for \textit{in-situ} annealing that may occur while being held in an optical trap at elevated temperature for extended periods of time up to several hours. Our results may be beneficial for improving the accuracy of force calibration in precision sensing experiments with optically trapped particles and for improving the trapping stability of nanoparticles in high vacuum conditions for a variety of levitated optomechanics experiments.

\section{Nanomechanical Testing} 

\begin{figure}
\begin{center}
\includegraphics[width=0.4\textwidth]{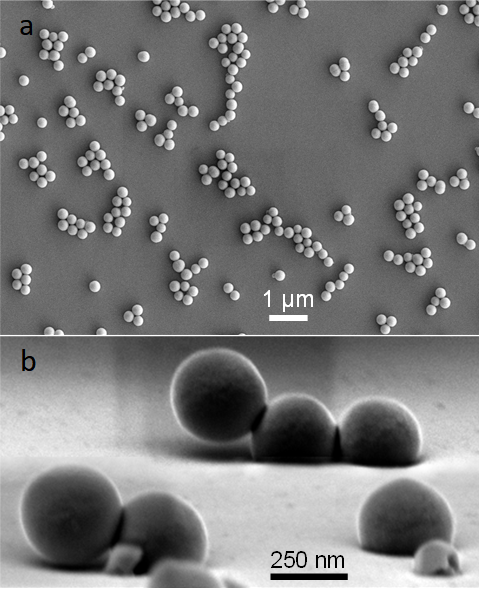}
\caption{\small{Scanning electron micrograph of (a) top-view and (b) side-view of spin-coated~300 nm diameter silica nanospheres on a silicon wafer, after annealing at $700^{\circ}$C for 1 hour in a N$_2$ atmosphere.\label{fig:spincoat}}}
\end{center}
\end{figure}

\emph{Sample Preparation--} Samples of silica nanospheres \fixme{available commercially from Bangs Laboratories Inc. Fishers, IN} were deposited by spin- coating methods on pieces of silicon wafers to produce a disperse uniform covering of the wafer (see Fig. 1). The spheres had a radius range of $280-305$ nm. The sizes of the wafer pieces were chosen to be approximately 1 cm x 1 cm, compatible with the furnace technology and SEM/nano-compression capabilities. Sets of spin-coated silica nanosphere samples were run in a flow furnace with nitrogen annealing under different conditions: $25^{\circ}$C (room temperature, $298$ K, $0.15$ $T_m$), $300^{\circ}$C ($576$ K, $0.30$ $T_m$) for 1 hour, $450^{\circ}$C ($723$ K, $0.37$ $T_m$) for 1 hour, and $700^{\circ}$C ($976$ K, $0.50$ $T_m$) for 1 hour, where $T_m$ is the homologous temperature (melting point of silica in K).

To assess the quality of the annealed nanosphere samples and to ensure that they are free of contaminants, we first examined them using a scanning electron microscope (SEM) prior to nano-compression experiments. 
To prepare them for SEM they are sputter coated with 5 nm of platinum in order to prevent the samples from being charged under the electron beam. 
Energy Dispersive Spectroscopy (EDS) was employed to verify the chemical composition of the \fixme{nanospheres} and verify that those samples used in the nanomechanical testing were relatively free of gross impurities, including salts of Na and Ca. The data of the clean non-contaminated nanospheres showed that their atomic makeup is consistent with the manufacturer specifications.


\emph{Nano-compression experiments--}
\begin{figure}[!t]
\begin{center}
\includegraphics[width=0.48\textwidth]{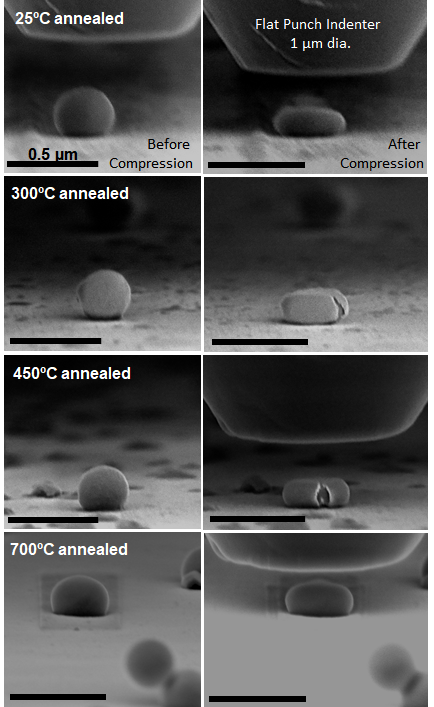}
\caption{\small{\label{fig:strainimage}} Silica nanospheres before and after compression to high strain levels of $0.65 -  0.75$ for four different annealing conditions. These tests were conducted in the ‘e-beam on’ condition.  
}
\end{center}
\end{figure}
 \textit{In-situ} SEM nano-compression testing was conducted in a nanomechanical instrument, which is comprised of a nano-mechanical tester (Hysitron PI-85) inside of an SEM (FEI Magellan and FEI Scios). The nanospheres were compressed with a flat punch conductive diamond tip of 1 $\mu$m diameter, at a nominal displacement rate of 2 nm/s. The continuously captured image scans were recorded as a video file during the test (see Supporting Information). For each annealing temperature, the silica nanospheres were compressed to two different strain levels. Four nanospheres were compressed to lower strain levels of $0.25-0.35$, while another four were compressed to higher strain levels of $0.65-0.75$ (see Fig. \ref{fig:loadtest}).

However, at the nano/micro length scales the mechanical properties of silica are known to be severely affected by the electron beam, and such e-beam induced plasticity has been reported to result in a four-fold difference in the measured flow stresses even when the ambient temperatures are near room temperatures \cite{Zheng2010}. In order to account for any such superplastic flow in our silica nanospheres, a parallel set of experiments were conducted in which the e-beam was turned off during the nano-compression tests (i.e. after the flat punch was in position on top of the nanosphere). Both sets of experiments were analyzed using the protocols described below.   

The elastic moduli of the nanospheres were calculated from the measured load-displacement data for both the high and low strain levels using Eqs. 1-6 below. 

A spherical shape model developed by Lin et al. \cite{YL2008} was used to calculate the contact radii $a$ of the spheres using Equation (\ref{eq1}):
\begin{equation}
a=\sqrt{R^{\prime2}-{(R-\delta)}^2}    					   \label{eq1}
\end{equation}
where $2\delta$ is the total displacement in compression, $R$ is the initial radius of the sphere, and $R'$ is the diagonal radius of the compressed sphere (see Fig. \ref{fig:loadtest}) which can be obtained using Eq. \ref{eq2}:  

\begin{figure}[!t]
\begin{center}
\includegraphics[width=0.5\textwidth]{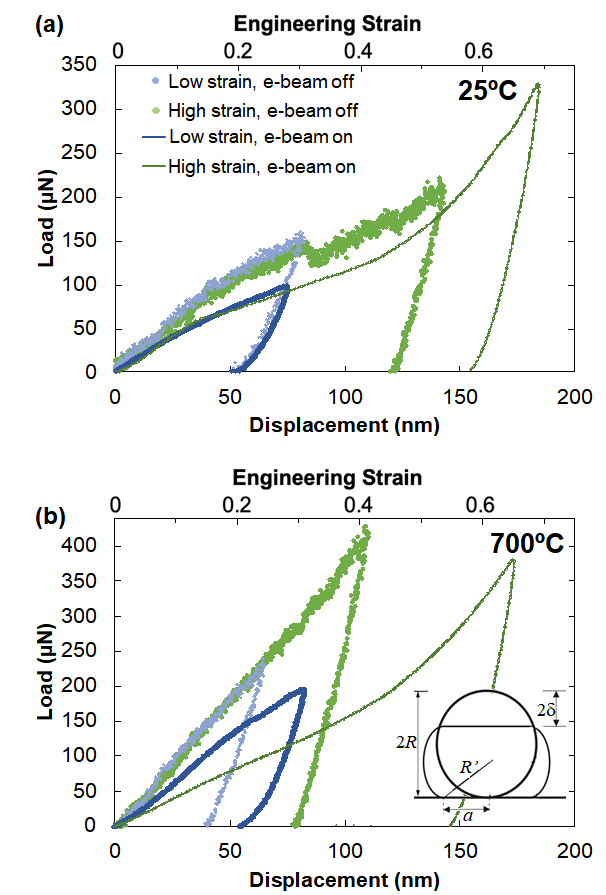}
\caption{\small{\label{fig:loadtest}} Representative load-displacement responses for Si nanospheres at high and low strain levels with the electron beam on and off at (a) 25$^\circ$C (298 K, 0.15 T$_m$) and (b) after annealing at 700$^\circ$C (976 K, 0.50 T$_m$) for 1 hour. (b inset) Schematic diagram of a compressed nanosphere under mechanical load. Adapted from Ref. \cite{YL2008}. 
}
\end{center}
\end{figure}

\begin{equation}
R^{\prime2}=\frac{2R^3}{3(R-\delta)}+\frac{{(R-\delta)}^2}{3}.  \label{eq2} \end{equation}
The values of $R'$ from Eq. \ref{eq2} were also cross-checked using the SEM measurements taken during the \textit{in-situ} tests. 


Additionally, we applied Sneddon's compliance correction (Eq. (\ref{eq3})) to the measured stiffness values in order to account for the elastic displacement of the substrate during compression \cite{SNEDDON196547}.

\begin{equation}C_{\mathrm{Sneddon}}=\frac{\sqrt\pi(1-v^2)}{2E_s\sqrt{A_s}}    	\label{eq3}			   
\end{equation}

Here, $E_s$ is the elastic modulus of silica, $A_s$ is the instantaneous contact area of the nanosphere, and $v=0.0361$ is the Poisson ratio of silica. The true stiffness $S_{\mathrm{corrected}}$ is then calculated by subtracting the Sneddon compliance ($C_{\mathrm{Sneddon}}$) from the inverse of the stiffness measured during the nano-compression experiment ($S_{\mathrm{measured}}$) as
\begin{equation}
S_{\mathrm{corrected}}=\frac{1}{\left(\frac{1}{S_{\mathrm{meas}}}\right)-C_{\mathrm{Sneddon}}}.    \label{eq4}				  
\end{equation}

Finally, the values of the sample moduli were computed following Hertz theory \cite{Hertz} as:
\begin{eqnarray}
E_{\mathrm{eff}}&=&\frac{S_{\mathrm{corrected}}}{2a}    					   
\\
\frac{1}{E_{\mathrm{eff}}}&=&\frac{1-v_s^2}{E_s}+\frac{1-v_i^2}{E_i}    	 \end{eqnarray}

Here $E_{\mathrm{eff}}$ is the effective stiffness of the indenter and the specimen system, $v$ and $E$ are the Poisson’s ratio and the Young’s modulus, and the subscripts $s$ and $i$ refer to the specimen and the indenter, respectively ($E_i = 1140$ GPa, $v_i = 0.07$). $S_{\mathrm{measured}} = \frac{dP}{d(2\delta)}$, is the experimentally measured stiffness of the upper 50\% of the unloading data, and $P$ is the measured load. Additional values of $S_{\mathrm{measured}}$ were also calculated using 30\%, and 70\% of the unloading curve for comparison. These results are described in Table \ref{table1} 

\begin{figure}[!t]
\begin{center}
\includegraphics[width=0.5\textwidth]{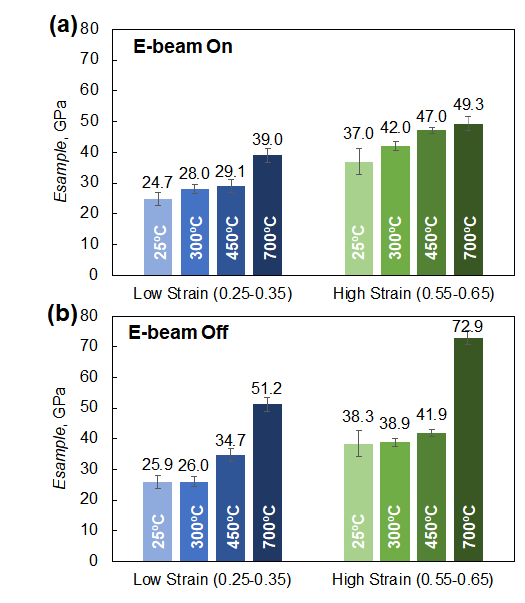}
\caption{\small{\label{fig:lowhigh}} Comparison of the measured elastic modulus values for the silica nanospheres annealed at varying temperatures, then compressed to low (0.25-0.35) and high (0.55-0.65) strain levels for tests with (a) the electron beam on and (b) beam off conditions.}
\end{center}
\end{figure}

\emph{Results from nano-compression experiments--} The results from the nano-compression experiments on the silica nanospheres are shown in Figs. \ref{fig:strainimage}, \ref{fig:loadtest} and \ref{fig:lowhigh} and in Table \ref{table1}. Fig. \ref{fig:strainimage} shows the silica nanospheres before and after compression to a high strain level of $0.65 - 0.75$ (also see the video files in Supporting Information). The images in this figure are from tests conducted with the electron beam on throughout the experiment. It is interesting to note that most of the nanospheres, such as the ones shown in Fig. \ref{fig:strainimage} that were annealed at $300^{\circ}$C and $450^{\circ}$C,  show a crack forming during the nano-compression process. The initiation of such a crack can also often be traced to a discontinuity in their respective load-displacement responses, and occurs primarily when the nanospheres are compressed to the high strain levels of $0.65 - 0.75$ (see Fig. \ref{fig:loadtest}  and the video files in Supporting Information section). Most of the recorded load-displacement curves at these higher strain levels show such a discontinuity, suggesting that such cracks occur in almost all nanosphere samples (although some might be occurring on the back surface of the nanospheres and are not visible in the \textit{in-situ} videos). Tests at the lower strain levels of $0.25 - 0.35$ do not show any cracking in the SEM videos/images, nor do they show any discontinuities in their load-displacement responses. It is also interesting to note that these cracks occur in random local areas of the sphere volume, and they appear to ‘peel off’ the top layer of the nanosphere samples. This observation suggests that the regions below the outer top layer of the silica nanospheres might have a different density than the rest of the material, and this density difference is non-uniform across the nanosphere volume.       

Figs. \ref{fig:loadtest}a and \ref{fig:loadtest}b shows representative load-displacement (strain) nano-compression responses of the silica nanosphere samples annealed at room temperature ($25^{\circ}$C) and $700^{\circ}$C respectively. A combination of responses from both low and high strain levels, and tests with the e-beam on vs. off are shown. A primary observation is that the e-beam introduces significant plasticity in the nanosphere. For example, at a representative strain of $0.4$ for the room temperature samples, the load in the e-beam off condition is $60$ \% higher than when the e-beam is kept on during the experiment ($200$ vs. $125$ $\mu$N). This is equivalent of a stress increase of 8.5\% ($1.65$ vs. $1.52$ GPa). The difference is even higher for the silica nanospheres annealed at $700^{\circ}$C. For these samples at a representative strain of $0.4$, the load is $143$\% higher ($425$ vs. $175$ $\mu$N) and the stress is $60$\% higher ($4.63$ vs. $3.75$ GPa) for the e-beam off vs. on conditions. Hence the mechanical properties for the experiments conducted with the ‘e-beam on’ are expected to be significantly affected (i.e. lower) than their actual values. 


Overall the load-displacement curves show a convex shape with the slope decreasing as load (or displacement) increases. This is typical of nano-compression experiments on spherical particles where the contact area increases with load \cite{YL2008}. The curves show a sharper increase in their slopes at strains $>0.4$, which is presumably when densification of the material is initiated under compression especially under the 'e-beam on' condition. Such a slope increase due to densification is reminiscent of the behavior of porous structures such as vertically aligned carbon nanotube (VACNT) bundles \cite{PATHAK2009,Pathak2013} and open cell foams \cite{gibson_ashby_1997}, further indicating that the as-deposited silica nanospheres are of a lower-than-full density.



Figure 6 summarizes the elastic modulus computed from all nano-compression experiments, for both e-beam on (Fig. \ref{fig:lowhigh}a) and e-beam off (Fig. \ref{fig:lowhigh}b) conditions, and at different strain levels for all four annealing conditions. First, we note that the tests conducted without keeping the e-beam on shows higher modulus values for all conditions, as discussed earlier. Similarly, as expected, the tests stopped before the onset of densification at the low strain ($0.25-0.35$) levels show a lower modulus as compared to the tests at the higher strain ($0.55-0.65$) levels. The highest modulus value of $72.9\pm2.3$ GPa is obtained from the nano-compression tests on silica nanospheres annealed at $700^{\circ}$C ($976$ K, $0.50$Tm, Fig. \ref{fig:lowhigh}b), which is identical to the modulus of bulk silica \cite{Deschamps2014}. 

Assuming spherical voids, the approximate porosity $p$ in the silica nanospheres can be calculated from the values of the measured $E$ vs. ideal $E_0$ modulus using MacKinzie's equation \cite{Mackenzie_1950}
\begin{equation}
E = E_0 (1 - 1.9p). \label{eq7}
\end{equation}
If $E$ and $E_0$ are taken to be the measured modulus at high strain for the $25$ and $700^{\circ}$C tests respectively (Fig. \ref{fig:lowhigh}b and Table \ref{table1}), then Equation \ref{eq7} indicates that the silica nanospheres have an approximate porosity of $25$\%. In comparison, the manufacturer reports a $\sim 10$\% porosity (calculated from the density provided by the manufacturer, 2 g/cm$^3$ vs. the bulk density of silica 2.2 g/cm$^3$). We expect the actual porosity of the nanospheres to lie within this range; i.e. within 10-25\%.  

We note that the modulus values shown in Fig. \ref{fig:lowhigh} are dependent on the measured stiffness, which is typically calculated using the upper 50\% of the unloading data. Changing the extent of the unloading slope for modulus calculations from 50\% to 30\% or 70\% causes the measured sample moduli to increase or decrease respectively, as shown in Table \ref{table1}. The data in Table \ref{table1} is shown primarily for completeness of our approach; the insights described above remain unchanged even though the absolute modulus values can vary when using different extents of the unloading slopes.   

Another point of discussion is the wettability of the silica nanospheres on the glass substrate as a function of the annealing temperatures. At the highest annealing temperature ($700^{\circ}$C), the nanospheres are observed to wet the substrate considerably, and correspondingly appear less spherical. Figure \ref{fig:strainimage} shows the difference in sphericity of the nanospheres for the four different annealing temperatures, where the nanospheres annealed at $700^{\circ}$C are noticeably less spherical. Notably, the nanospheres that are not in contact with the substrate but are instead only touching other nanospheres retain a more spherical shape (see Fig. \ref{fig:spincoat}b and Fig. \ref{fig:strainimage} bottom images). This is similar to other observations in literature, such as platinum nanocrystals on amorphous silica films \cite{Yu2005}, where annealing temperatures of higher than $500^{\circ}$C can cause an increased wetting of the two contact surfaces. 

Although annealing at $700^{\circ}$C produces near bulk-silica-like behavior, possibly leading to improved trapping stability in levitated optomechanics experiments, a change in sphericity of the nanospheres at these temperatures can have significant implications on their trapping stability. To investigate the impact of annealing on optical trapping stability, we conducted additional optical trapping experiments with annealed and as-grown nanospheres. These results, and a theoretical analysis of related trap stability considerations, are described in Sec. \ref{sec:trapping} below.


\section{Optical Trapping Tests\label{sec:trapping}}
\begin{figure}
\begin{center}
\includegraphics[width=0.45\textwidth]{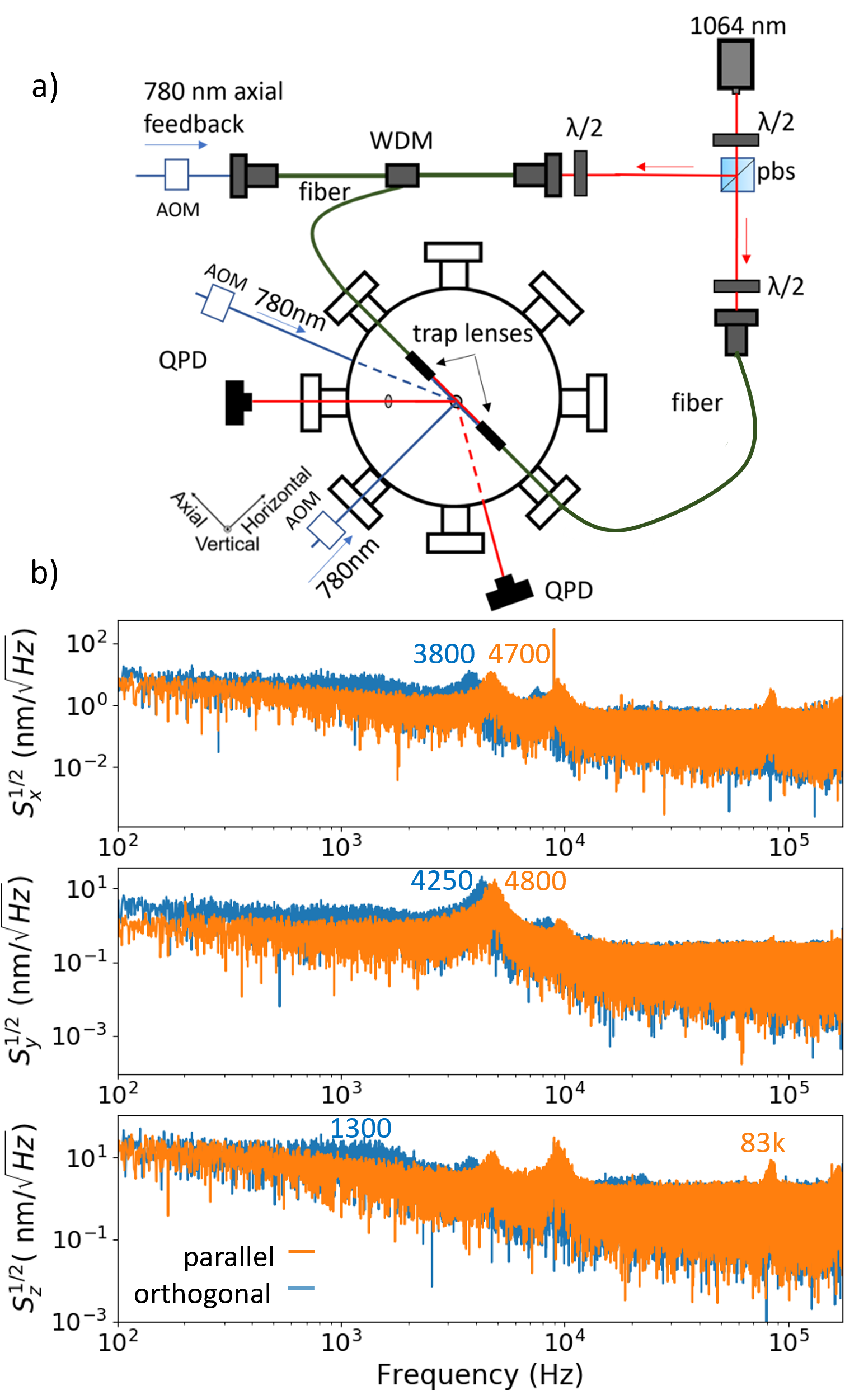}
\caption{\small{\label{fig:expdiag}}  a) Schematic of experimental setup used for testing the trapping stability for nanoparticles in a dual-beam optical dipole trap in high vacuum. b) Typical displacement spectral density of a $300$ nm sphere in the dual-beam fiber-coupled optical trap. For orthogonal polarization of the dual beams (blue), the frequencies along the horizontal ($S_x$), vertical ($S_y$) and axial ($S_z$) directions are about 3800 Hz, 4250 Hz, and 1300 Hz respectively. Parallel polarization (orange) increases the frequencies to 4700 Hz, 4800 Hz, and 83 kHz. The change in the horizontal frequency implies the intensity increased by at most $1.5 \times$, indicating imperfect interference.}
\end{center}
\end{figure}

\emph{Experimental setup.--} The stability of annealed and as-grown $300$ nm silica spheres in optical traps was tested in the experimental setup depicted in Fig. \ref{fig:expdiag}.  In the setup a 1064nm laser is split into two paths by a polarizing beam splitter and fiber coupled into the vacuum chamber. The fiber output lenses inside the chamber are mounted at a separation of $34$ mm (approximately 2$\times$ their focal length) with a foci separation of about 75$\mu$m. Fine anti-parallel alignment and foci separation are set by a 3-axis kinematic mount holding one of the fiber output lenses. Both lenses are affixed to a platform which is attached to a stack of 3 piezo actuated translation stages oriented along the axial, vertical and horizontal axes. The trap is formed by the balance of scattering forces of the two counter-propagating beams focused down to waists of $6.5$ $\mu$m. The sphere trapped at the center of the dual beam foci can then be positioned with micron precision along each axis in the trap.  In addition, half-wave plates before the input into polarization maintaining fibers allow for the polarization of the two beams to be rotated to maximize their interference and the intensity on the trapped sphere. \fixme{Scattered light from trapped nanospheres is imaged onto two quadrant photodetectors (QPDs), viewing the particle from the side and the bottom of the chamber, respectively, to determine the 3-dimensional displacement spectral density of the motion of the trapped particles.} Fig. \ref{fig:expdiag}(b) shows a typical 3-dimensional spectrum of the nanosphere’s motion when the trap is in the orthogonal and parallel polarization configurations. Ideally the trap intensity increases by a factor of 2 upon polarization rotation, resulting in a frequency shift in the transverse (horizontal and vertical) motion of the bead by a factor of $\sqrt{2}$. The axial frequency increases dramatically due to the large intensity gradient of the standing wave. 

To maintain the trapped sphere while pumping to high vacuum, 780 nm beams modulated by acousto-optic modulators (AOMs) with active feedback were used to cool the sphere’s center of mass motion along the three axes shown in Fig. \ref{fig:expdiag}, using similar methods to those used in our prior work \cite{ranjit2016}. \fixme{In order to provide cooling along three axes, a wave-division multiplexer (WDM) is used to combine the 780 nm cooling light and 1064 nm trapping light into one of the two fibers for cooling along the axial (beam) direction, and two additional 780 nm cooling beams are directed from the upwards and horizontal transverse directions, respectively.} Once at pressures of $<10^{-5}$mbar the residual gas is insufficient to cool the surface of the sphere and it reaches a high equilibrium temperature determined by a balance between emitted and absorbed blackbody radiation and power absorbed by the trapping laser. Trap stability at high vacuum was evaluated by increasing the intensity of the laser in the maximally interfering trap configuration and recording when the sphere was lost from the trap.

\begin{figure}[H]
\begin{center}
\includegraphics[width=0.48\textwidth]{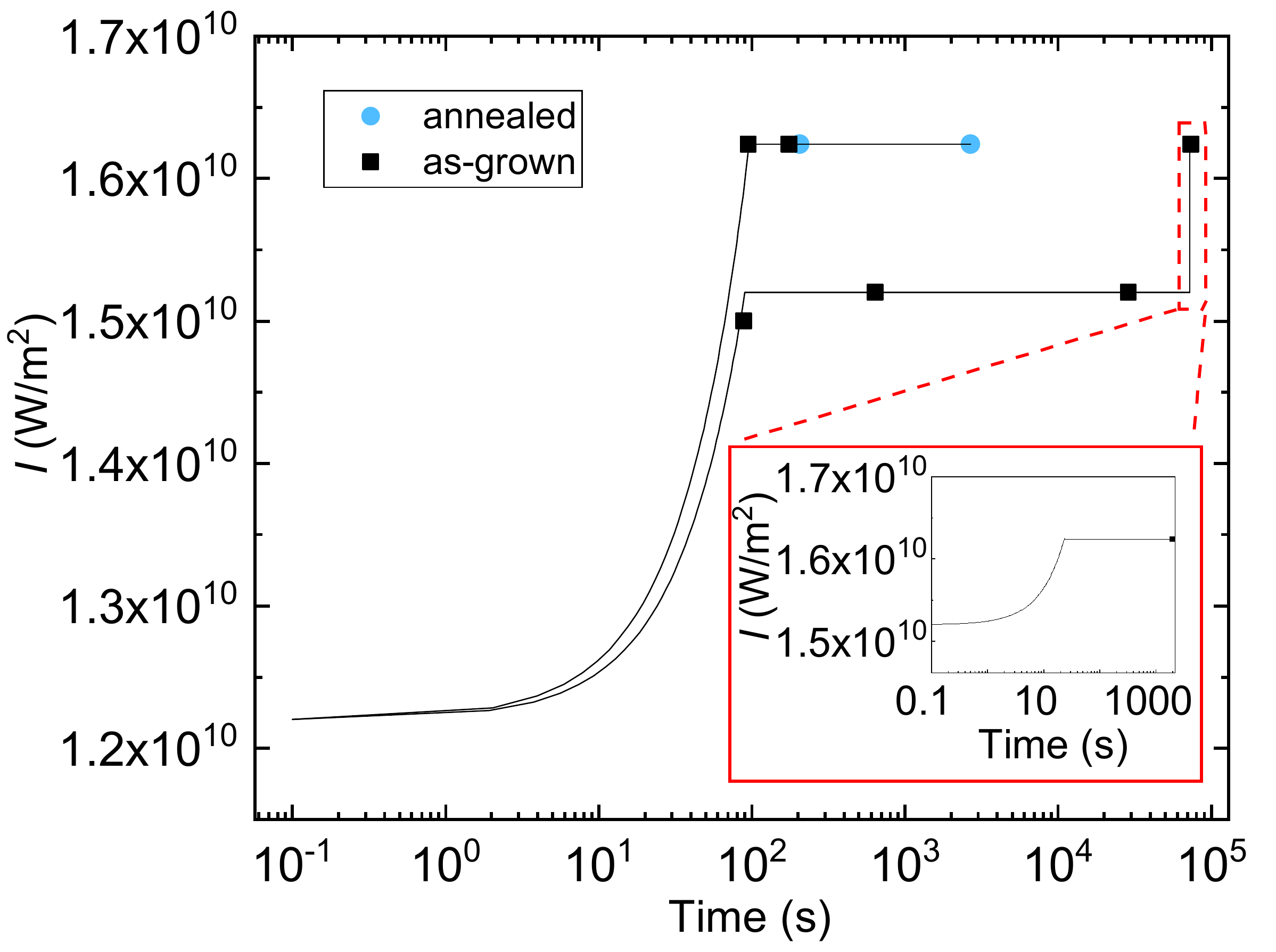}
\caption{\small{\label{fig:IvT}} Trap intensity versus time as-grown and annealed 300nm silica spheres survived at high vacuum. Data points indicate when the sphere was lost along its heating path. We note a reduced trapping rate of annealed particles compared to as-grown, possibly as a result of the particles becoming much less spherical during annealing at $700^{\circ}$C as mentioned in the following sections.}
\end{center}
\end{figure}

\emph{Trap stability tests.--} Results for the intensity versus trapping lifetime at high vacuum for as-grown and annealed spheres (at $700^{\circ}$C) are shown in Fig. \ref{fig:IvT}. The spheres were initially trapped, cooled and then pumped to high vacuum at a trap intensity of 1.22$\times10^{10}$ W/m$^2$. Once the experimental chamber reached $\sim$ 10$^{-6}$ mbar, the trap intensity was then gradually increased to 1.52$\times10^{10}$W/m$^2$ or 1.62$\times10^{10}$ W/m$^2$, 
where we note that spheres along the higher intensity path are lost more quickly. 

Our prior results have suggested that the plastic properties of the silica nanospheres are severely affected by the e-beam, which can cause e-beam induced plasticity (Figs. \ref{fig:loadtest} and \ref{fig:lowhigh}). Similarly, heating under the high intensity trapping laser can also cause in-situ annealing of the nanospheres, which can be significant for our experimental parameters. For example, the Inset of Fig. \ref{fig:IvT} shows that one particular as-grown sphere resided at 1.52$\times10^{10}$W/m$^2$ for about 74000 seconds before the trap intensity was again increased to 1.62$\times10^{10}$ W/m$^2$ where it survived an additional 3000 seconds. At such high vacuum, cooling from background gas collisions becomes less significant and the equilibrium surface temperature of the nanosphere is determined by the balance of power absorbed by the laser and blackbody emission and absorption from the surrounding environment \cite{ranjit2016}. Prior simulations indicate that for such trap intensities, for example an equilibrium internal temperature of $400^{\circ}$C-$700^{\circ}$C is predicted for an imaginary component of the dielectric permittivity $\epsilon_2$ ranging from $\epsilon_2 = 2.5 \times 10^{-7}$ (corresponding to bulk silica glass) to $\epsilon_2 = 1.0 \times 10^6$ \cite{ranjit2016}.  Although these results warrant further study and more data, the current data is suggestive in that annealing may help increase the duration a sphere remains trapped at high intensity, and also suggests the possibility that nanospheres can undergo annealing while trapped at high intensity for a prolonged time. 

\emph{Analysis of Trapping stability--}
\fixme{The results from the nano-compression experiments indicate that (i) the as-deposited silica nanospheres likely have considerable (10-25\%) porosity, (ii) this porosity is likely localized and not uniformly distributed across the nanosphere volume and (iii) while annealing can help with improving the properties of the nanospheres to a more bulk-like response, it might also cause them to become non-spherical in shape. In order to quantify these effects, we consider how modification of the density, size, mass, or refractive index of the nanosphere impacts the optical forces acting on it and the trap stability. In particular, we calculate the scattering and gradient forces on the particle under various scenarios, and finally consider the momentum imparted on the trapped nanoparticle should mass be ejected from the particle while it is being heated by the trapping laser.} 

\fixme{Changing the radius to accommodate a density change 
leads to quantitatively different optical forces. However, if the density change occurs uniformly and the particle retains a spherical shape, the optical forces remain balanced, and thus the smaller sphere remains stably trapped.}
\begin{figure}[!t]
\begin{center}
\includegraphics[width=0.5\textwidth]{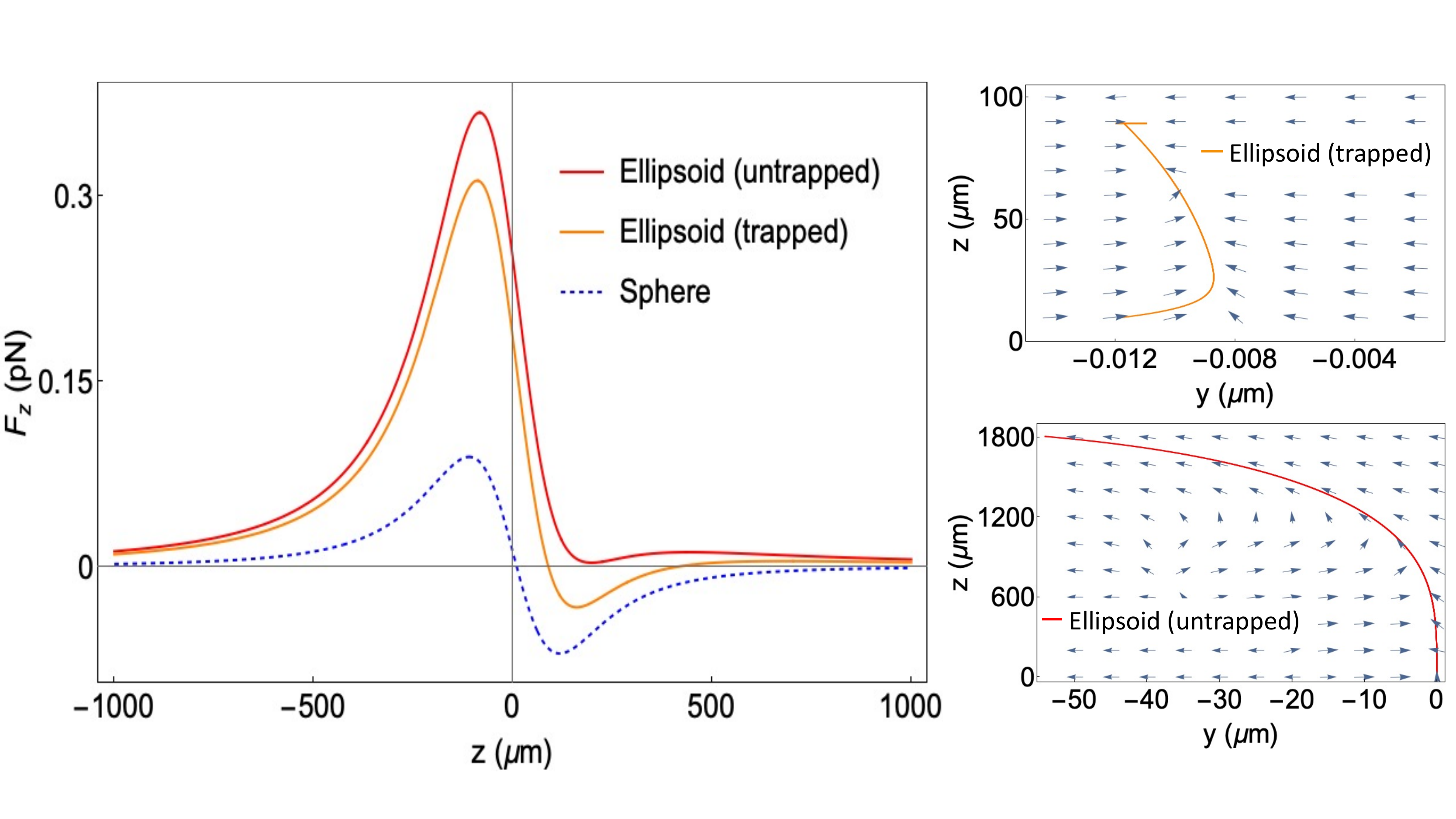}
\caption{\small{\label{fig:ellipsoid}} (Left:) Optical forces on a levitated particle in the axial direction. The three plots correspond to the following scenarios: a sphere of radius 150nm, density $\rho = 2$g/cm$^3$ and refractive index $n_p$ = 1.46 (blue); an ellipsoid with $\delta r = 20$nm, $\rho = 2.2$ g/cm$^3$ and $n_p$= 1.8 (orange); and an ellipsoid with $\delta r = 28$nm, $\rho = 2.2$ g/cm$^3$ and $n_p$= 1.8 (red). (Right:) Vector plot of the optical forces and corresponding particle trajectories for the two ellipsoidal particles in consideration.}
\end{center}
\end{figure}
The optical forces are no longer balanced if the shrinking of the particle is not symmetric, for example if the particle changes from spherical to ellipsoidal. We calculate that such a shape change occuring while the particle is trapped would cause the particle to get kicked out of the trap for sufficient deformation, with a low asymmetry threshold for particle loss for the case of higher density. Below this threshold, the particle remains trapped, albeit at a slightly different location. Fig. \ref{fig:ellipsoid} shows the optical forces in these scenarios, along with the particle trajectories, showing eventual particle loss. The method for analytically evaluating the optical forces for sphere and ellipsoid are described in the Appendix, and elsewhere \cite{Harada1996, Trojek2012} and the experimental parameters are the same as those from Ref. \cite{ranjit2016}.

For the trap stability analysis, we assume that the trapped particle is a sphere of radius $a_s= 150$nm and density $2.0$ g/cm$^3$ before annealing. We assume that as a result of the annealing process, the shape of the particle changes from sphere to ellipsoid with the longest axis along the $x$ axis (semi-axis $a$), and shortest axis along the $y$ axis (semi-axis $b$). The third axis is adjusted to ensure conservation of mass. This asymmetric shape change produces the maximum imbalance in the scattering forces along $z$ direction that would have otherwise canceled each other for a sphere due to equal power in the two counter-propagating beams. 

Keeping all the material properties constant, we find that a fractional change of radius ($\delta r = (a-a_s)/a_s= (a_s-b)/a_s$) of 0.3 is required for the axial forces to not be balanced at any $z$. Such an imbalance of forces would kick the particle out of the trap. For smaller changes in radii, the forces balance for a non-zero positive $z$ value, indicating a shift of equilibrium position. This is confirmed by mapping a particle's trajectory in the presence of such forces, as shown in inset of Fig. \ref{fig:ellipsoid}.

The transition to an non-trappable shape happens for smaller asymmetry for slightly different material and optical trapping conditions. For example, a density change from 2 to 2.2g/cm$^3$, accompanied by an increase of 5\% in the refractive index $n_p$ and a power imbalance of 1\%, enables the particle to be lost at a much smaller $\delta r$ of $0.25$. \fixme{Given the degree of asymmetry required for destabilizing the trap, we conclude that such a shape change alone is unlikely to explain the observed particle loss. 
However spheres that are sufficiently deformed with an ellipsoidal shape while stuck to the trap loading substrate, e.g. as shown in Fig. \ref{fig:spincoat}b may be impossible to load into the optical trap to begin with.}

In addition to this study, we numerically studied the effect of having non-concentric spot/hole in the sphere with a different index (taking $n=1$), to simulate the effect of having a trapped  air or water bubble. In the Rayleigh regime, the effect of such an asymmetry is to rotate the scatter pattern. Realistic scenarios lead to under a 1 degree rotation, and thus would not be sufficient to account for particle loss. 

Finally, we investigated the effect of mass ejection, for example if a small water bubble embedded in the nanosphere migrated to the surface and evaporated under the sustained heating by the optical trap. We find that a mass loss of 0.022\% at $20^\circ$C provides the nanosphere with enough momentum for it to escape the trap, assuming that the particle has an initial speed due to thermal motion and the water bubble leaves at the root-mean-squared speed of water molecules. At an elevated temperature of $700^\circ$C, a mass loss of only 0.010\% would be sufficient to kick the particle out of the trap. We consider this to be a more likely mechanism of the observed sudden particle loss from the optical trap.

\section{Conclusion}

Our results have shown that the density and elastic modulus of the silica nanospheres increases with annealing temperature due to densification. After annealing at $700^{\circ}$C 
the elastic modulus is approaching that of fused silica, while for lesser temperatures the nanospheres are still porous. We have studied the trap lifetime of annealed and as-grown silica nanospheres in an optical trap in high vacuum, and find trap instabilities at high laser intensity in both samples, with the annealed samples indicating slightly improved lifetimes. We identify several possible mechanisms for trap loss, including a change in the particle density, shape, refractive index, and size that may occur while the particle is suspended in the laser trap at elevated temperature. Although other mechanisms remain a possibility, we find trapping instabilities under high vacuum at high trap laser intensity may be most likely explained by momentum-recoil from ejection of trapped water or air from the nanoparticles after being heated for extended time periods in the optical trap. Achieving bulk-like silica behavior while retaining the spherical shape of the nanoparticle may lead to improved trapping stability in future levitated optomechanics experiments. 


\begin{table*}[!t]
\begin{tabular}{|l|l|l|l|l|l|l|l|l|}
\hline
\multicolumn{9}{|l|}{(a)~E$_{\text{sample}}$ (GPa), e-beam on}                                                                                                                                                                                                                                                                      \\ \hline
                                  & \multicolumn{2}{l|}{25ºC}                                         & \multicolumn{2}{l|}{300ºC}                                        & \multicolumn{2}{l|}{450ºC}                                        & \multicolumn{2}{l|}{700ºC}                                        \\
                                  & \multicolumn{2}{l|}{(298 K, 0.15Tm)}                              & \multicolumn{2}{l|}{(576 K, 0.30Tm)}                              & \multicolumn{2}{l|}{(723 K, 0.37Tm)}                              & \multicolumn{2}{l|}{(976 K, 0.50Tm)}                              \\ \cline{2-9} 
{Annealing temp.} & {\color[HTML]{3166FF} LS}       & {\color[HTML]{009901} HS}       & {\color[HTML]{3166FF} LS}       & {\color[HTML]{009901} HS}       & {\color[HTML]{3166FF} LS}       & {\color[HTML]{009901} HS}       & {\color[HTML]{3166FF} LS}       & {\color[HTML]{009901} HS}       \\ \hline
30\% Unload                       & {\color[HTML]{3166FF} $28.8 \pm 6.3$} & {\color[HTML]{009901} $40.8\pm2.1$} & {\color[HTML]{3166FF} $36.5\pm1.1$} & {\color[HTML]{009901} $50.4\pm3.0$} & {\color[HTML]{3166FF} $36.3\pm5.0$} & {\color[HTML]{009901} $57.1\pm7.9$} & {\color[HTML]{3166FF} $47.5\pm1.9$} & {\color[HTML]{009901} $55.9\pm9.4$} \\ \hline
50\% Unload                       & {\color[HTML]{3166FF} $24.7\pm4.5$} & {\color[HTML]{009901} $37.0\pm1.8$} & {\color[HTML]{3166FF} $36.5\pm2.0$} & {\color[HTML]{009901} $42.0\pm0.8$} & {\color[HTML]{3166FF} $29.1\pm3.1$} & {\color[HTML]{009901} $47.0\pm7.4$} & {\color[HTML]{3166FF} $39.0\pm1.2$} & {\color[HTML]{009901} $49.3\pm5.9$} \\ \hline
70\% Unload                       & {\color[HTML]{3166FF} $21.6\pm4.3$} & {\color[HTML]{009901} $33.8\pm1.3$} & {\color[HTML]{3166FF} $23.7\pm2.7$} & {\color[HTML]{009901} $35.5\pm1.5$} & {\color[HTML]{3166FF} $24.9\pm3.0$} & {\color[HTML]{009901} $40.1\pm6.9$} & {\color[HTML]{3166FF} $34.1\pm1.4$} & {\color[HTML]{009901} $44.4\pm4.9$} \\ \hline
\multicolumn{9}{|l|}{(b)~E$_{\text{sample}}$ (GPa), e-beam off}                                                                                                                                                                                                                                                                      \\ \hline
                                  & \multicolumn{2}{l|}{25ºC}                                         & \multicolumn{2}{l|}{300ºC}                                        & \multicolumn{2}{l|}{450ºC}                                        & \multicolumn{2}{l|}{700ºC}                                        \\
                                  & \multicolumn{2}{l|}{(298 K, 0.15Tm)}                              & \multicolumn{2}{l|}{(576 K, 0.30Tm)}                              & \multicolumn{2}{l|}{(723 K, 0.37Tm)}                              & \multicolumn{2}{l|}{(976 K, 0.50Tm)}                              \\ \cline{2-9} 
{Annealing temp.} & {\color[HTML]{3166FF} LS}       & {\color[HTML]{009901} HS}       & {\color[HTML]{3166FF} LS}       & {\color[HTML]{009901} HS}       & {\color[HTML]{3166FF} LS}       & {\color[HTML]{009901} HS}       & {\color[HTML]{3166FF} LS}       & {\color[HTML]{009901} HS}       \\ \hline
30\% Unload                       & {\color[HTML]{3166FF} $30.4\pm4.3$} & {\color[HTML]{009901} $41.1\pm2.2$} & {\color[HTML]{3166FF} $27.9\pm3.0$} & {\color[HTML]{009901} $40.6\pm3.3$} & {\color[HTML]{3166FF} $36.3\pm2.3$} & {\color[HTML]{009901} $43.1\pm2.4$} & {\color[HTML]{3166FF} $53.7\pm2.4$} & {\color[HTML]{009901} $72.7\pm2.9$} \\ \hline
50\% Unload                       & {\color[HTML]{3166FF} $25.9\pm1.4$} & {\color[HTML]{009901} $38.3\pm1.6$} & {\color[HTML]{3166FF} $26.0\pm1.8$} & {\color[HTML]{009901} $38.9\pm1.8$} & {\color[HTML]{3166FF} $34.4\pm3.3$} & {\color[HTML]{009901} $41.8\pm3.7$} & {\color[HTML]{3166FF} $51.2\pm2.3$} & {\color[HTML]{009901} $72.9\pm2.3$} \\ \hline
70\% Unload                       & {\color[HTML]{3166FF} $24.3\pm1.1$} & {\color[HTML]{009901} $34.5\pm2.2$} & {\color[HTML]{3166FF} $24.8\pm2.4$} & {\color[HTML]{009901} $36.9\pm2.2$} & {\color[HTML]{3166FF} $34.2\pm3.2$} & {\color[HTML]{009901} $40.5\pm3.2$} & {\color[HTML]{3166FF} $49.6\pm2.7$} & {\color[HTML]{009901} $68.3\pm3.7$} \\ \hline
\end{tabular}
\caption{\small{\label{table1}} Average ± standard deviation elastic modulus values calculated using varying extents (30, 50 and 70\%) of the unloading slope for silica nanosphere samples annealed at varying temperatures and then compressed to low strain (LS = 0.25-0.35) and high strain (HS=0.55-0.65) levels for tests under (a) electron beam on and (b) off conditions. 
}
\end{table*}

\section{Appendix}
In Table \ref{table1} we report the values and uncertainties corresponding to the elastic moduli shown in Fig. \ref{fig:lowhigh} as determined using two strain rates and varying percentages of the unload segment in the load-displacement curves.

We also include formulas used to estimate the optical scattering rates for spherical and elliptical particles in the Rayleigh limit, which should be a reasonable approximation for the optical trapping parameters considered.

\subsection{Optical trapping of ellipsoidal particles}
This section presents the details of the optical trapping stability analysis. The average force exerted by an arbitrary time-harmonic electromagnetic field $\Vec{E}$ on a small particle (with dimensions $\ll$ wavelength of light,) is given by \cite{Chaumet2000}
\begin{equation}
F_j = \frac{1}{2} \Re{\{\Vec{p}^* \cdot \partial_j \Vec{E}\}},
\end{equation}
 where $F_j$ denotes the component of (scattering + gradient) force along the $j-$ direction, and $\Vec{p}=\hat{\alpha} \Vec{E}$ is the dipole moment of the particle with $\hat{\alpha}$ being the polarizability tensor.
 
 We consider an ellipsoidal particle with semiaxes $a,b,c$ along the $x,y,z$ directions respectively. The component of polarizability  along the $j$-axis is given by \cite{Trojek2012}
\begin{equation}
    \alpha_j = \alpha'_j + i\alpha''_j = \frac{\alpha_j^0}{1-\frac{i k^3 \alpha_j^0}{6 \pi \epsilon_m}},
\end{equation}
where
\begin{equation}
\alpha_j^0 = \frac{\epsilon_0 n_m^2 V}{L_j + \frac{1}{m^2-1}}, 
\end{equation}
$m = n_p/n_m$, and
\begin{equation}
    L_x = \int_0^\infty \frac{a b c \text{d} s}{2 (s+a^2)^{3/2} (s+b^2)^{1/2} (s+c^2)^{1/2}}.
    \label{L_j}
\end{equation}
$L_y$ and $L_z$ can be found by exchanging the semi-axes in the denominator in Eq.~(\ref{L_j}). Here, $n_p$ is the refractive index of the particle, $n_m$ is the refractive index of the medium, and volume $V = 4\pi abc/3$. For our simulations, we assume $n_p=1.46$ and $n_m$=1.

Along the principal axes coordinate system, the polarizability tensor has the form
\begin{equation}
    \hat{\alpha} = \begin{bmatrix}
\alpha'_x+i\alpha''_x & 0 & 0\\
0 & \alpha'_y+i\alpha''_y & 0\\
0 & 0 & \alpha'_z+i\alpha''_z
\end{bmatrix}.
\end{equation}
The polarizability along a rotated coordinate system can be evaluated by applying the appropriate rotational transformation \cite{Trojek2012}. Since we are concerned with estimating the change in optical forces as a result of deformation from sphere to ellipsoid, we assume that the optical axes are aligned with the principal axes of the ellipsoid in this work. 

We calculate optical forces in the dual beam configuration discussed in Ref.~\cite{ranjit2016} and Section 2. We consider the first Gaussian beam to be propagating along the $+z$ direction and polarized along the $+x$ direction. The second beam is considered to be polarized along the $+y$ direction and traveling along the $-z$ direction. The beam foci are offset by 75 $\mu$m, and the total optical power of the two beams is 2.2W. 

In our simulations, the trapped particle is initially a sphere of radius $a_s= 150$ nm and density $2.0$ g/cm$^3$. It undergoes the annealing process and becomes ellipsoidal with the longest axis along the $x$ direction (semi-axis $a$), and the shortest axis along the $y$ direction (semi-axis $b$). The semi-axis $c$ also changes so that mass is conserved. Our goal is to find the fractional change of radius $\delta r =  (a-a_s)/a_s =  (a_s-b)/a_s$ required for the optical forces along the $z$ direction to not be balanced at any $z$.

Assuming that all the material properties remain constant throughout the annealing process, we find that a fractional radius change of 0.3 produces sufficient imbalance in the axial forces to kick the particle out of the trap. By changing the density and refractive index of the particle, we are able to see particle loss for even smaller asymmetry. For example, an increase in $n_p$ of 5\% accompanied by a density change from 2 to 2.2g/cm$^3$ would cause the particle to not be trapped for $\delta r= 0.28$. Similar refractive index modification due to density change has been observed at microwave frequencies \cite{Hotta2009}. Along with material properties, the optical stability geometry is highly sensitive to power imbalance in the trapping beams. For example, a power imbalance of even 1\% in the previous scenario changes $\delta r$ to $0.25$. 
Nevertheless, the shape change required to transition into an unstable geometry for optical trapping is quite significant. Therefore, we conclude that a shape change alone is unlikely to explain particle loss. 

\section{Acknowledgements}
We thank G. Winstone for useful discussions and assistance with numerical simulations of Rayleigh scattering. AG is partially supported by NSF grants PHY-1805994, PHY-2110524 and the Heising-Simons Foundation, the W.M. Keck Foundation, the J. Templeton Foundation, and the office of Naval Research grant no.417315//N00014-18-1-2370. Instrumentation funding for the nano-compression experiments was provided through the NSF MRI no. 1726897 (SP and AG) and DURIP no. W911NF2110042 (SP) grant. SS would like to acknowledge support from NSF grants PHY-1912480 and PHY-2047707. This work was performed, in part, at the Center for Integrated Nanotechnologies, an Office of Science User Facility operated for the U.S. Department of Energy (DOE) Office of Science by Los Alamos National Laboratory (Contract 89233218CNA000001) and Sandia National Laboratories (Contract DE-NA-0003525).    

\section{Supplemental Material Information}
The continuously captured images from the in-situ scanning electron microscopy (SEM) nano-compression experiments under the e-beam on condition were recorded as video files. The four video files in the Supplemental Material section show the nanocompression response of a representative silica nanosphere from each of the four annealing conditions (25, 200, 450 and 700 ºC).     

\bibliography{library-apd,library_jw,library_geraci} 

\begin{thebibliography}{43}%
\makeatletter
\providecommand \@ifxundefined [1]{%
 \@ifx{#1\undefined}
}%
\providecommand \@ifnum [1]{%
 \ifnum #1\expandafter \@firstoftwo
 \else \expandafter \@secondoftwo
 \fi
}%
\providecommand \@ifx [1]{%
 \ifx #1\expandafter \@firstoftwo
 \else \expandafter \@secondoftwo
 \fi
}%
\providecommand \natexlab [1]{#1}%
\providecommand \enquote  [1]{``#1''}%
\providecommand \bibnamefont  [1]{#1}%
\providecommand \bibfnamefont [1]{#1}%
\providecommand \citenamefont [1]{#1}%
\providecommand \href@noop [0]{\@secondoftwo}%
\providecommand \href [0]{\begingroup \@sanitize@url \@href}%
\providecommand \@href[1]{\@@startlink{#1}\@@href}%
\providecommand \@@href[1]{\endgroup#1\@@endlink}%
\providecommand \@sanitize@url [0]{\catcode `\\12\catcode `\$12\catcode
  `\&12\catcode `\#12\catcode `\^12\catcode `\_12\catcode `\%12\relax}%
\providecommand \@@startlink[1]{}%
\providecommand \@@endlink[0]{}%
\providecommand \url  [0]{\begingroup\@sanitize@url \@url }%
\providecommand \@url [1]{\endgroup\@href {#1}{\urlprefix }}%
\providecommand \urlprefix  [0]{URL }%
\providecommand \Eprint [0]{\href }%
\providecommand \doibase [0]{http://dx.doi.org/}%
\providecommand \selectlanguage [0]{\@gobble}%
\providecommand \bibinfo  [0]{\@secondoftwo}%
\providecommand \bibfield  [0]{\@secondoftwo}%
\providecommand \translation [1]{[#1]}%
\providecommand \BibitemOpen [0]{}%
\providecommand \bibitemStop [0]{}%
\providecommand \bibitemNoStop [0]{.\EOS\space}%
\providecommand \EOS [0]{\spacefactor3000\relax}%
\providecommand \BibitemShut  [1]{\csname bibitem#1\endcsname}%
\let\auto@bib@innerbib\@empty
\bibitem [{\citenamefont {Gieseler}\ \emph {et~al.}(2012)\citenamefont
  {Gieseler}, \citenamefont {Deutsch}, \citenamefont {Quidant},\ and\
  \citenamefont {Novotny}}]{novotny2012}%
  \BibitemOpen
  \bibfield  {author} {\bibinfo {author} {\bibfnamefont {J.}~\bibnamefont
  {Gieseler}}, \bibinfo {author} {\bibfnamefont {B.}~\bibnamefont {Deutsch}},
  \bibinfo {author} {\bibfnamefont {R.}~\bibnamefont {Quidant}}, \ and\
  \bibinfo {author} {\bibfnamefont {L.}~\bibnamefont {Novotny}},\ }\href
  {\doibase 10.1103/PhysRevLett.109.103603} {\bibfield  {journal} {\bibinfo
  {journal} {Phys. Rev. Lett.}\ }\textbf {\bibinfo {volume} {109}},\ \bibinfo
  {pages} {103603} (\bibinfo {year} {2012})}\BibitemShut {NoStop}%
\bibitem [{\citenamefont {Ranjit}\ \emph {et~al.}(2016)\citenamefont {Ranjit},
  \citenamefont {Cunningham}, \citenamefont {Casey},\ and\ \citenamefont
  {Geraci}}]{ranjit2016}%
  \BibitemOpen
  \bibfield  {author} {\bibinfo {author} {\bibfnamefont {G.}~\bibnamefont
  {Ranjit}}, \bibinfo {author} {\bibfnamefont {M.}~\bibnamefont {Cunningham}},
  \bibinfo {author} {\bibfnamefont {K.}~\bibnamefont {Casey}}, \ and\ \bibinfo
  {author} {\bibfnamefont {A.~A.}\ \bibnamefont {Geraci}},\ }\href {\doibase
  10.1103/PhysRevA.93.053801} {\bibfield  {journal} {\bibinfo  {journal} {Phys.
  Rev. A}\ }\textbf {\bibinfo {volume} {93}},\ \bibinfo {pages} {053801}
  (\bibinfo {year} {2016})}\BibitemShut {NoStop}%
\bibitem [{\citenamefont {Geraci}\ \emph {et~al.}(2010)\citenamefont {Geraci},
  \citenamefont {Papp},\ and\ \citenamefont {Kitching}}]{geraci2010}%
  \BibitemOpen
  \bibfield  {author} {\bibinfo {author} {\bibfnamefont {A.~A.}\ \bibnamefont
  {Geraci}}, \bibinfo {author} {\bibfnamefont {S.~B.}\ \bibnamefont {Papp}}, \
  and\ \bibinfo {author} {\bibfnamefont {J.}~\bibnamefont {Kitching}},\ }\href
  {\doibase 10.1103/PhysRevLett.105.101101} {\bibfield  {journal} {\bibinfo
  {journal} {Phys. Rev. Lett.}\ }\textbf {\bibinfo {volume} {105}},\ \bibinfo
  {pages} {101101} (\bibinfo {year} {2010})}\BibitemShut {NoStop}%
\bibitem [{\citenamefont {Geraci}\ and\ \citenamefont
  {Goldman}(2015)}]{andyhart2015}%
  \BibitemOpen
  \bibfield  {author} {\bibinfo {author} {\bibfnamefont {A.}~\bibnamefont
  {Geraci}}\ and\ \bibinfo {author} {\bibfnamefont {H.}~\bibnamefont
  {Goldman}},\ }\href {\doibase 10.1103/PhysRevD.92.062002} {\bibfield
  {journal} {\bibinfo  {journal} {Phys. Rev. D}\ }\textbf {\bibinfo {volume}
  {92}},\ \bibinfo {pages} {062002} (\bibinfo {year} {2015})}\BibitemShut
  {NoStop}%
\bibitem [{\citenamefont {Monteiro}\ \emph {et~al.}(2017)\citenamefont
  {Monteiro}, \citenamefont {Ghosh}, \citenamefont {Fine},\ and\ \citenamefont
  {Moore}}]{Moore2017}%
  \BibitemOpen
  \bibfield  {author} {\bibinfo {author} {\bibfnamefont {F.}~\bibnamefont
  {Monteiro}}, \bibinfo {author} {\bibfnamefont {S.}~\bibnamefont {Ghosh}},
  \bibinfo {author} {\bibfnamefont {A.~G.}\ \bibnamefont {Fine}}, \ and\
  \bibinfo {author} {\bibfnamefont {D.~C.}\ \bibnamefont {Moore}},\ }\href
  {\doibase 10.1103/PhysRevA.96.063841} {\bibfield  {journal} {\bibinfo
  {journal} {Phys. Rev. A}\ }\textbf {\bibinfo {volume} {96}},\ \bibinfo
  {pages} {063841} (\bibinfo {year} {2017})}\BibitemShut {NoStop}%
\bibitem [{\citenamefont {Hebestreit}\ \emph {et~al.}(2018)\citenamefont
  {Hebestreit}, \citenamefont {Frimmer}, \citenamefont {Reimann},\ and\
  \citenamefont {Novotny}}]{novotnydrop}%
  \BibitemOpen
  \bibfield  {author} {\bibinfo {author} {\bibfnamefont {E.}~\bibnamefont
  {Hebestreit}}, \bibinfo {author} {\bibfnamefont {M.}~\bibnamefont {Frimmer}},
  \bibinfo {author} {\bibfnamefont {R.}~\bibnamefont {Reimann}}, \ and\
  \bibinfo {author} {\bibfnamefont {L.}~\bibnamefont {Novotny}},\ }\href
  {\doibase 10.1103/PhysRevLett.121.063602} {\bibfield  {journal} {\bibinfo
  {journal} {Phys. Rev. Lett.}\ }\textbf {\bibinfo {volume} {121}},\ \bibinfo
  {pages} {063602} (\bibinfo {year} {2018})}\BibitemShut {NoStop}%
\bibitem [{\citenamefont {Hoang}\ \emph {et~al.}(2016)\citenamefont {Hoang},
  \citenamefont {Ma}, \citenamefont {Ahn}, \citenamefont {Bang}, \citenamefont
  {Robicheaux}, \citenamefont {Yin},\ and\ \citenamefont {Li}}]{Li2016}%
  \BibitemOpen
  \bibfield  {author} {\bibinfo {author} {\bibfnamefont {T.~M.}\ \bibnamefont
  {Hoang}}, \bibinfo {author} {\bibfnamefont {Y.}~\bibnamefont {Ma}}, \bibinfo
  {author} {\bibfnamefont {J.}~\bibnamefont {Ahn}}, \bibinfo {author}
  {\bibfnamefont {J.}~\bibnamefont {Bang}}, \bibinfo {author} {\bibfnamefont
  {F.}~\bibnamefont {Robicheaux}}, \bibinfo {author} {\bibfnamefont {Z.-Q.}\
  \bibnamefont {Yin}}, \ and\ \bibinfo {author} {\bibfnamefont
  {T.}~\bibnamefont {Li}},\ }\href {\doibase 10.1103/PhysRevLett.117.123604}
  {\bibfield  {journal} {\bibinfo  {journal} {Phys. Rev. Lett.}\ }\textbf
  {\bibinfo {volume} {117}},\ \bibinfo {pages} {123604} (\bibinfo {year}
  {2016})}\BibitemShut {NoStop}%
\bibitem [{\citenamefont {Ahn}\ \emph {et~al.}(2018)\citenamefont {Ahn},
  \citenamefont {Xu}, \citenamefont {Bang}, \citenamefont {Deng}, \citenamefont
  {Hoang}, \citenamefont {Han}, \citenamefont {Ma},\ and\ \citenamefont
  {Li}}]{Li2018}%
  \BibitemOpen
  \bibfield  {author} {\bibinfo {author} {\bibfnamefont {J.}~\bibnamefont
  {Ahn}}, \bibinfo {author} {\bibfnamefont {Z.}~\bibnamefont {Xu}}, \bibinfo
  {author} {\bibfnamefont {J.}~\bibnamefont {Bang}}, \bibinfo {author}
  {\bibfnamefont {Y.-H.}\ \bibnamefont {Deng}}, \bibinfo {author}
  {\bibfnamefont {T.~M.}\ \bibnamefont {Hoang}}, \bibinfo {author}
  {\bibfnamefont {Q.}~\bibnamefont {Han}}, \bibinfo {author} {\bibfnamefont
  {R.-M.}\ \bibnamefont {Ma}}, \ and\ \bibinfo {author} {\bibfnamefont
  {T.}~\bibnamefont {Li}},\ }\href {\doibase 10.1103/PhysRevLett.121.033603}
  {\bibfield  {journal} {\bibinfo  {journal} {Phys. Rev. Lett.}\ }\textbf
  {\bibinfo {volume} {121}},\ \bibinfo {pages} {033603} (\bibinfo {year}
  {2018})}\BibitemShut {NoStop}%
\bibitem [{\citenamefont {Reimann}\ \emph {et~al.}(2018)\citenamefont
  {Reimann}, \citenamefont {Doderer}, \citenamefont {Hebestreit}, \citenamefont
  {Diehl}, \citenamefont {Frimmer}, \citenamefont {Windey}, \citenamefont
  {Tebbenjohanns},\ and\ \citenamefont {Novotny}}]{Novotny2018}%
  \BibitemOpen
  \bibfield  {author} {\bibinfo {author} {\bibfnamefont {R.}~\bibnamefont
  {Reimann}}, \bibinfo {author} {\bibfnamefont {M.}~\bibnamefont {Doderer}},
  \bibinfo {author} {\bibfnamefont {E.}~\bibnamefont {Hebestreit}}, \bibinfo
  {author} {\bibfnamefont {R.}~\bibnamefont {Diehl}}, \bibinfo {author}
  {\bibfnamefont {M.}~\bibnamefont {Frimmer}}, \bibinfo {author} {\bibfnamefont
  {D.}~\bibnamefont {Windey}}, \bibinfo {author} {\bibfnamefont
  {F.}~\bibnamefont {Tebbenjohanns}}, \ and\ \bibinfo {author} {\bibfnamefont
  {L.}~\bibnamefont {Novotny}},\ }\href {\doibase
  10.1103/PhysRevLett.121.033602} {\bibfield  {journal} {\bibinfo  {journal}
  {Phys. Rev. Lett.}\ }\textbf {\bibinfo {volume} {121}},\ \bibinfo {pages}
  {033602} (\bibinfo {year} {2018})}\BibitemShut {NoStop}%
\bibitem [{\citenamefont {Monteiro}\ \emph {et~al.}(2018)\citenamefont
  {Monteiro}, \citenamefont {Ghosh}, \citenamefont {van Assendelft},\ and\
  \citenamefont {Moore}}]{Moore2018}%
  \BibitemOpen
  \bibfield  {author} {\bibinfo {author} {\bibfnamefont {F.}~\bibnamefont
  {Monteiro}}, \bibinfo {author} {\bibfnamefont {S.}~\bibnamefont {Ghosh}},
  \bibinfo {author} {\bibfnamefont {E.~C.}\ \bibnamefont {van Assendelft}}, \
  and\ \bibinfo {author} {\bibfnamefont {D.~C.}\ \bibnamefont {Moore}},\ }\href
  {\doibase 10.1103/PhysRevA.97.051802} {\bibfield  {journal} {\bibinfo
  {journal} {Phys. Rev. A}\ }\textbf {\bibinfo {volume} {97}},\ \bibinfo
  {pages} {051802} (\bibinfo {year} {2018})}\BibitemShut {NoStop}%
\bibitem [{\citenamefont {Romero-Isart}\ \emph {et~al.}(2011)\citenamefont
  {Romero-Isart}, \citenamefont {Pflanzer}, \citenamefont {Juan}, \citenamefont
  {Quidant}, \citenamefont {Kiesel}, \citenamefont {Aspelmeyer},\ and\
  \citenamefont {Cirac}}]{oriol2011}%
  \BibitemOpen
  \bibfield  {author} {\bibinfo {author} {\bibfnamefont {O.}~\bibnamefont
  {Romero-Isart}}, \bibinfo {author} {\bibfnamefont {A.~C.}\ \bibnamefont
  {Pflanzer}}, \bibinfo {author} {\bibfnamefont {M.~L.}\ \bibnamefont {Juan}},
  \bibinfo {author} {\bibfnamefont {R.}~\bibnamefont {Quidant}}, \bibinfo
  {author} {\bibfnamefont {N.}~\bibnamefont {Kiesel}}, \bibinfo {author}
  {\bibfnamefont {M.}~\bibnamefont {Aspelmeyer}}, \ and\ \bibinfo {author}
  {\bibfnamefont {J.~I.}\ \bibnamefont {Cirac}},\ }\href {\doibase
  10.1103/PhysRevA.83.013803} {\bibfield  {journal} {\bibinfo  {journal} {Phys.
  Rev. A}\ }\textbf {\bibinfo {volume} {83}},\ \bibinfo {pages} {013803}
  (\bibinfo {year} {2011})}\BibitemShut {NoStop}%
\bibitem [{\citenamefont {Chang}\ \emph {et~al.}(2010)\citenamefont {Chang},
  \citenamefont {Regal}, \citenamefont {Papp}, \citenamefont {Wilson},
  \citenamefont {Ye}, \citenamefont {Painter}, \citenamefont {Kimble},\ and\
  \citenamefont {Zoller}}]{chang2009}%
  \BibitemOpen
  \bibfield  {author} {\bibinfo {author} {\bibfnamefont {D.~E.}\ \bibnamefont
  {Chang}}, \bibinfo {author} {\bibfnamefont {C.~A.}\ \bibnamefont {Regal}},
  \bibinfo {author} {\bibfnamefont {S.~B.}\ \bibnamefont {Papp}}, \bibinfo
  {author} {\bibfnamefont {D.~J.}\ \bibnamefont {Wilson}}, \bibinfo {author}
  {\bibfnamefont {J.}~\bibnamefont {Ye}}, \bibinfo {author} {\bibfnamefont
  {O.}~\bibnamefont {Painter}}, \bibinfo {author} {\bibfnamefont {H.~J.}\
  \bibnamefont {Kimble}}, \ and\ \bibinfo {author} {\bibfnamefont
  {P.}~\bibnamefont {Zoller}},\ }\href {\doibase 10.1073/pnas.0912969107}
  {\bibfield  {journal} {\bibinfo  {journal} {Proceedings of the National
  Academy of Sciences}\ }\textbf {\bibinfo {volume} {107}},\ \bibinfo {pages}
  {1005} (\bibinfo {year} {2010})},\ \Eprint
  {http://arxiv.org/abs/https://www.pnas.org/content/107/3/1005.full.pdf}
  {https://www.pnas.org/content/107/3/1005.full.pdf} \BibitemShut {NoStop}%
\bibitem [{\citenamefont {Windey}\ \emph {et~al.}(2019)\citenamefont {Windey},
  \citenamefont {Gonzalez-Ballestero}, \citenamefont {Maurer}, \citenamefont
  {Novotny}, \citenamefont {Romero-Isart},\ and\ \citenamefont
  {Reimann}}]{coherentscattering}%
  \BibitemOpen
  \bibfield  {author} {\bibinfo {author} {\bibfnamefont {D.}~\bibnamefont
  {Windey}}, \bibinfo {author} {\bibfnamefont {C.}~\bibnamefont
  {Gonzalez-Ballestero}}, \bibinfo {author} {\bibfnamefont {P.}~\bibnamefont
  {Maurer}}, \bibinfo {author} {\bibfnamefont {L.}~\bibnamefont {Novotny}},
  \bibinfo {author} {\bibfnamefont {O.}~\bibnamefont {Romero-Isart}}, \ and\
  \bibinfo {author} {\bibfnamefont {R.}~\bibnamefont {Reimann}},\ }\href
  {\doibase 10.1103/PhysRevLett.122.123601} {\bibfield  {journal} {\bibinfo
  {journal} {Phys. Rev. Lett.}\ }\textbf {\bibinfo {volume} {122}},\ \bibinfo
  {pages} {123601} (\bibinfo {year} {2019})}\BibitemShut {NoStop}%
\bibitem [{\citenamefont {Deli\ifmmode~\acute{c}\else \'{c}\fi{}}\ \emph
  {et~al.}(2019)\citenamefont {Deli\ifmmode~\acute{c}\else \'{c}\fi{}},
  \citenamefont {Reisenbauer}, \citenamefont {Grass}, \citenamefont {Kiesel},
  \citenamefont {Vuleti\ifmmode~\acute{c}\else \'{c}\fi{}},\ and\ \citenamefont
  {Aspelmeyer}}]{aspelmeyercavity}%
  \BibitemOpen
  \bibfield  {author} {\bibinfo {author} {\bibfnamefont {U.~c.~v.}\
  \bibnamefont {Deli\ifmmode~\acute{c}\else \'{c}\fi{}}}, \bibinfo {author}
  {\bibfnamefont {M.}~\bibnamefont {Reisenbauer}}, \bibinfo {author}
  {\bibfnamefont {D.}~\bibnamefont {Grass}}, \bibinfo {author} {\bibfnamefont
  {N.}~\bibnamefont {Kiesel}}, \bibinfo {author} {\bibfnamefont
  {V.}~\bibnamefont {Vuleti\ifmmode~\acute{c}\else \'{c}\fi{}}}, \ and\
  \bibinfo {author} {\bibfnamefont {M.}~\bibnamefont {Aspelmeyer}},\ }\href
  {\doibase 10.1103/PhysRevLett.122.123602} {\bibfield  {journal} {\bibinfo
  {journal} {Phys. Rev. Lett.}\ }\textbf {\bibinfo {volume} {122}},\ \bibinfo
  {pages} {123602} (\bibinfo {year} {2019})}\BibitemShut {NoStop}%
\bibitem [{\citenamefont {Ranjit}\ \emph {et~al.}(2015)\citenamefont {Ranjit},
  \citenamefont {Montoya},\ and\ \citenamefont {Geraci}}]{sympcool}%
  \BibitemOpen
  \bibfield  {author} {\bibinfo {author} {\bibfnamefont {G.}~\bibnamefont
  {Ranjit}}, \bibinfo {author} {\bibfnamefont {C.}~\bibnamefont {Montoya}}, \
  and\ \bibinfo {author} {\bibfnamefont {A.~A.}\ \bibnamefont {Geraci}},\
  }\href {\doibase 10.1103/PhysRevA.91.013416} {\bibfield  {journal} {\bibinfo
  {journal} {Phys. Rev. A}\ }\textbf {\bibinfo {volume} {91}},\ \bibinfo
  {pages} {013416} (\bibinfo {year} {2015})}\BibitemShut {NoStop}%
\bibitem [{\citenamefont {Moore}\ \emph {et~al.}(2014)\citenamefont {Moore},
  \citenamefont {Rider},\ and\ \citenamefont {Gratta}}]{millicharge}%
  \BibitemOpen
  \bibfield  {author} {\bibinfo {author} {\bibfnamefont {D.~C.}\ \bibnamefont
  {Moore}}, \bibinfo {author} {\bibfnamefont {A.~D.}\ \bibnamefont {Rider}}, \
  and\ \bibinfo {author} {\bibfnamefont {G.}~\bibnamefont {Gratta}},\ }\href
  {\doibase 10.1103/PhysRevLett.113.251801} {\bibfield  {journal} {\bibinfo
  {journal} {Phys. Rev. Lett.}\ }\textbf {\bibinfo {volume} {113}},\ \bibinfo
  {pages} {251801} (\bibinfo {year} {2014})}\BibitemShut {NoStop}%
\bibitem [{\citenamefont {Arvanitaki}\ and\ \citenamefont
  {Geraci}(2013)}]{GWprl}%
  \BibitemOpen
  \bibfield  {author} {\bibinfo {author} {\bibfnamefont {A.}~\bibnamefont
  {Arvanitaki}}\ and\ \bibinfo {author} {\bibfnamefont {A.~A.}\ \bibnamefont
  {Geraci}},\ }\href {\doibase 10.1103/PhysRevLett.110.071105} {\bibfield
  {journal} {\bibinfo  {journal} {Phys. Rev. Lett.}\ }\textbf {\bibinfo
  {volume} {110}},\ \bibinfo {pages} {071105} (\bibinfo {year}
  {2013})}\BibitemShut {NoStop}%
\bibitem [{\citenamefont {Aggarwal}\ \emph {et~al.}(2020)\citenamefont
  {Aggarwal}, \citenamefont {Winstone}, \citenamefont {Teo}, \citenamefont
  {Baryakhtar}, \citenamefont {Larson}, \citenamefont {Kalogera},\ and\
  \citenamefont {Geraci}}]{LSDpaper}%
  \BibitemOpen
  \bibfield  {author} {\bibinfo {author} {\bibfnamefont {N.}~\bibnamefont
  {Aggarwal}}, \bibinfo {author} {\bibfnamefont {G.~P.}\ \bibnamefont
  {Winstone}}, \bibinfo {author} {\bibfnamefont {M.}~\bibnamefont {Teo}},
  \bibinfo {author} {\bibfnamefont {M.}~\bibnamefont {Baryakhtar}}, \bibinfo
  {author} {\bibfnamefont {S.~L.}\ \bibnamefont {Larson}}, \bibinfo {author}
  {\bibfnamefont {V.}~\bibnamefont {Kalogera}}, \ and\ \bibinfo {author}
  {\bibfnamefont {A.~A.}\ \bibnamefont {Geraci}},\ }\href@noop {} {\  (\bibinfo
  {year} {2020})},\ \Eprint {http://arxiv.org/abs/2010.13157} {arXiv:2010.13157
  [gr-qc]} \BibitemShut {NoStop}%
\bibitem [{\citenamefont {Carney}\ \emph {et~al.}(2019)\citenamefont {Carney},
  \citenamefont {Hook}, \citenamefont {Liu}, \citenamefont {Taylor},\ and\
  \citenamefont {Zhao}}]{DMcarney}%
  \BibitemOpen
  \bibfield  {author} {\bibinfo {author} {\bibfnamefont {D.}~\bibnamefont
  {Carney}}, \bibinfo {author} {\bibfnamefont {A.}~\bibnamefont {Hook}},
  \bibinfo {author} {\bibfnamefont {Z.}~\bibnamefont {Liu}}, \bibinfo {author}
  {\bibfnamefont {J.~M.}\ \bibnamefont {Taylor}}, \ and\ \bibinfo {author}
  {\bibfnamefont {Y.}~\bibnamefont {Zhao}},\ }\href@noop {} {\  (\bibinfo
  {year} {2019})},\ \Eprint {http://arxiv.org/abs/1908.04797} {arXiv:1908.04797
  [hep-ph]} \BibitemShut {NoStop}%
\bibitem [{\citenamefont {Carney}\ \emph {et~al.}(2021)\citenamefont {Carney},
  \citenamefont {Krnjaic}, \citenamefont {Moore}, \citenamefont {Regal},
  \citenamefont {Afek}, \citenamefont {Bhave}, \citenamefont {Brubaker},
  \citenamefont {Corbitt}, \citenamefont {Cripe}, \citenamefont {Crisosto}
  \emph {et~al.}}]{Carney2021}%
  \BibitemOpen
  \bibfield  {author} {\bibinfo {author} {\bibfnamefont {D.}~\bibnamefont
  {Carney}}, \bibinfo {author} {\bibfnamefont {G.}~\bibnamefont {Krnjaic}},
  \bibinfo {author} {\bibfnamefont {D.~C.}\ \bibnamefont {Moore}}, \bibinfo
  {author} {\bibfnamefont {C.~A.}\ \bibnamefont {Regal}}, \bibinfo {author}
  {\bibfnamefont {G.}~\bibnamefont {Afek}}, \bibinfo {author} {\bibfnamefont
  {S.}~\bibnamefont {Bhave}}, \bibinfo {author} {\bibfnamefont
  {B.}~\bibnamefont {Brubaker}}, \bibinfo {author} {\bibfnamefont
  {T.}~\bibnamefont {Corbitt}}, \bibinfo {author} {\bibfnamefont
  {J.}~\bibnamefont {Cripe}}, \bibinfo {author} {\bibfnamefont
  {N.}~\bibnamefont {Crisosto}},  \emph {et~al.},\ }\href@noop {} {\bibfield
  {journal} {\bibinfo  {journal} {Quantum Science and Technology}\ }\textbf
  {\bibinfo {volume} {6}},\ \bibinfo {pages} {024002} (\bibinfo {year}
  {2021})}\BibitemShut {NoStop}%
\bibitem [{\citenamefont {Rider}\ \emph {et~al.}(2016)\citenamefont {Rider},
  \citenamefont {Moore}, \citenamefont {Blakemore}, \citenamefont {Louis},
  \citenamefont {Lu},\ and\ \citenamefont {Gratta}}]{Rider2016}%
  \BibitemOpen
  \bibfield  {author} {\bibinfo {author} {\bibfnamefont {A.~D.}\ \bibnamefont
  {Rider}}, \bibinfo {author} {\bibfnamefont {D.~C.}\ \bibnamefont {Moore}},
  \bibinfo {author} {\bibfnamefont {C.~P.}\ \bibnamefont {Blakemore}}, \bibinfo
  {author} {\bibfnamefont {M.}~\bibnamefont {Louis}}, \bibinfo {author}
  {\bibfnamefont {M.}~\bibnamefont {Lu}}, \ and\ \bibinfo {author}
  {\bibfnamefont {G.}~\bibnamefont {Gratta}},\ }\href@noop {} {\bibfield
  {journal} {\bibinfo  {journal} {Physical review letters}\ }\textbf {\bibinfo
  {volume} {117}},\ \bibinfo {pages} {101101} (\bibinfo {year}
  {2016})}\BibitemShut {NoStop}%
\bibitem [{\citenamefont {Betz}\ \emph {et~al.}(2022)\citenamefont {Betz},
  \citenamefont {Manley}, \citenamefont {Wright}, \citenamefont {Grin},\ and\
  \citenamefont {Singh}}]{Betz2022}%
  \BibitemOpen
  \bibfield  {author} {\bibinfo {author} {\bibfnamefont {J.}~\bibnamefont
  {Betz}}, \bibinfo {author} {\bibfnamefont {J.}~\bibnamefont {Manley}},
  \bibinfo {author} {\bibfnamefont {E.}~\bibnamefont {Wright}}, \bibinfo
  {author} {\bibfnamefont {D.}~\bibnamefont {Grin}}, \ and\ \bibinfo {author}
  {\bibfnamefont {S.}~\bibnamefont {Singh}},\ }\href@noop {} {\bibfield
  {journal} {\bibinfo  {journal} {arXiv preprint arXiv:2201.12372}\ } (\bibinfo
  {year} {2022})}\BibitemShut {NoStop}%
\bibitem [{\citenamefont {{Dimopoulos}}\ and\ \citenamefont
  {{Giudice}}(1996)}]{GiudiceDimopoulos}%
  \BibitemOpen
  \bibfield  {author} {\bibinfo {author} {\bibfnamefont {S.}~\bibnamefont
  {{Dimopoulos}}}\ and\ \bibinfo {author} {\bibfnamefont {G.~F.}\ \bibnamefont
  {{Giudice}}},\ }\href {\doibase 10.1016/0370-2693(96)00390-5} {\bibfield
  {journal} {\bibinfo  {journal} {Physics Letters B}\ }\textbf {\bibinfo
  {volume} {379}},\ \bibinfo {pages} {105} (\bibinfo {year} {1996})},\ \Eprint
  {http://arxiv.org/abs/hep-ph/9602350} {hep-ph/9602350} \BibitemShut {NoStop}%
\bibitem [{\citenamefont {Arkani-Hamed}\ \emph {et~al.}(1999)\citenamefont
  {Arkani-Hamed}, \citenamefont {Dimopoulos},\ and\ \citenamefont
  {Dvali}}]{add}%
  \BibitemOpen
  \bibfield  {author} {\bibinfo {author} {\bibfnamefont {N.}~\bibnamefont
  {Arkani-Hamed}}, \bibinfo {author} {\bibfnamefont {S.}~\bibnamefont
  {Dimopoulos}}, \ and\ \bibinfo {author} {\bibfnamefont {G.}~\bibnamefont
  {Dvali}},\ }\href {\doibase 10.1103/PhysRevD.59.086004} {\bibfield  {journal}
  {\bibinfo  {journal} {Phys. Rev. D}\ }\textbf {\bibinfo {volume} {59}},\
  \bibinfo {pages} {086004} (\bibinfo {year} {1999})}\BibitemShut {NoStop}%
\bibitem [{\citenamefont {Kapner}\ \emph {et~al.}(2007)\citenamefont {Kapner},
  \citenamefont {Cook}, \citenamefont {Adelberger}, \citenamefont {Gundlach},
  \citenamefont {Heckel}, \citenamefont {Hoyle},\ and\ \citenamefont
  {Swanson}}]{Kapner2007}%
  \BibitemOpen
  \bibfield  {author} {\bibinfo {author} {\bibfnamefont {D.~J.}\ \bibnamefont
  {Kapner}}, \bibinfo {author} {\bibfnamefont {T.~S.}\ \bibnamefont {Cook}},
  \bibinfo {author} {\bibfnamefont {E.~G.}\ \bibnamefont {Adelberger}},
  \bibinfo {author} {\bibfnamefont {J.~H.}\ \bibnamefont {Gundlach}}, \bibinfo
  {author} {\bibfnamefont {B.~R.}\ \bibnamefont {Heckel}}, \bibinfo {author}
  {\bibfnamefont {C.~D.}\ \bibnamefont {Hoyle}}, \ and\ \bibinfo {author}
  {\bibfnamefont {H.~E.}\ \bibnamefont {Swanson}},\ }\href {\doibase
  10.1103/PhysRevLett.98.021101} {\bibfield  {journal} {\bibinfo  {journal}
  {Physical Review Letters}\ }\textbf {\bibinfo {volume} {98}},\ \bibinfo
  {pages} {021101} (\bibinfo {year} {2007})},\ \Eprint
  {http://arxiv.org/abs/0611184} {arXiv:0611184 [hep-ph]} \BibitemShut
  {NoStop}%
\bibitem [{\citenamefont {Geraci}\ \emph {et~al.}(2008)\citenamefont {Geraci},
  \citenamefont {Smullin}, \citenamefont {Weld}, \citenamefont {Chiaverini},\
  and\ \citenamefont {Kapitulnik}}]{Geraci2008}%
  \BibitemOpen
  \bibfield  {author} {\bibinfo {author} {\bibfnamefont {A.~A.}\ \bibnamefont
  {Geraci}}, \bibinfo {author} {\bibfnamefont {S.~J.}\ \bibnamefont {Smullin}},
  \bibinfo {author} {\bibfnamefont {D.~M.}\ \bibnamefont {Weld}}, \bibinfo
  {author} {\bibfnamefont {J.}~\bibnamefont {Chiaverini}}, \ and\ \bibinfo
  {author} {\bibfnamefont {A.}~\bibnamefont {Kapitulnik}},\ }\href {\doibase
  10.1103/PhysRevD.78.022002} {\bibfield  {journal} {\bibinfo  {journal} {Phys.
  Rev. D}\ }\textbf {\bibinfo {volume} {78}},\ \bibinfo {pages} {022002}
  (\bibinfo {year} {2008})}\BibitemShut {NoStop}%
\bibitem [{\citenamefont {Chen}\ \emph {et~al.}(2016)\citenamefont {Chen},
  \citenamefont {Tham}, \citenamefont {Krause}, \citenamefont {L\'opez},
  \citenamefont {Fischbach},\ and\ \citenamefont {Decca}}]{Chen2016}%
  \BibitemOpen
  \bibfield  {author} {\bibinfo {author} {\bibfnamefont {Y.-J.}\ \bibnamefont
  {Chen}}, \bibinfo {author} {\bibfnamefont {W.~K.}\ \bibnamefont {Tham}},
  \bibinfo {author} {\bibfnamefont {D.~E.}\ \bibnamefont {Krause}}, \bibinfo
  {author} {\bibfnamefont {D.}~\bibnamefont {L\'opez}}, \bibinfo {author}
  {\bibfnamefont {E.}~\bibnamefont {Fischbach}}, \ and\ \bibinfo {author}
  {\bibfnamefont {R.~S.}\ \bibnamefont {Decca}},\ }\href {\doibase
  10.1103/PhysRevLett.116.221102} {\bibfield  {journal} {\bibinfo  {journal}
  {Phys. Rev. Lett.}\ }\textbf {\bibinfo {volume} {116}},\ \bibinfo {pages}
  {221102} (\bibinfo {year} {2016})}\BibitemShut {NoStop}%
\bibitem [{\citenamefont {Kitamura}\ \emph {et~al.}(2007)\citenamefont
  {Kitamura}, \citenamefont {Pilon},\ and\ \citenamefont
  {Jonasz}}]{silicaglass}%
  \BibitemOpen
  \bibfield  {author} {\bibinfo {author} {\bibfnamefont {R.}~\bibnamefont
  {Kitamura}}, \bibinfo {author} {\bibfnamefont {L.}~\bibnamefont {Pilon}}, \
  and\ \bibinfo {author} {\bibfnamefont {M.}~\bibnamefont {Jonasz}},\ }\href
  {\doibase 10.1364/AO.46.008118} {\bibfield  {journal} {\bibinfo  {journal}
  {Appl. Opt.}\ }\textbf {\bibinfo {volume} {46}},\ \bibinfo {pages} {8118}
  (\bibinfo {year} {2007})}\BibitemShut {NoStop}%
\bibitem [{\citenamefont {BrÃ¼ckner}(1970)}]{bruckner}%
  \BibitemOpen
  \bibfield  {author} {\bibinfo {author} {\bibfnamefont {R.}~\bibnamefont
  {BrÃ¼ckner}},\ }\href {\doibase
  https://doi.org/10.1016/0022-3093(70)90190-0} {\bibfield  {journal} {\bibinfo
   {journal} {Journal of Non-Crystalline Solids}\ }\textbf {\bibinfo {volume}
  {5}},\ \bibinfo {pages} {123 } (\bibinfo {year} {1970})}\BibitemShut
  {NoStop}%
\bibitem [{\citenamefont {Zheng}\ \emph {et~al.}(2010)\citenamefont {Zheng},
  \citenamefont {Wang}, \citenamefont {Cheng}, \citenamefont {Yue},
  \citenamefont {Han}, \citenamefont {Zhang}, \citenamefont {Shan},
  \citenamefont {Mao}, \citenamefont {Ye}, \citenamefont {Yin},\ and\
  \citenamefont {Ma}}]{Zheng2010}%
  \BibitemOpen
  \bibfield  {author} {\bibinfo {author} {\bibfnamefont {K.}~\bibnamefont
  {Zheng}}, \bibinfo {author} {\bibfnamefont {C.}~\bibnamefont {Wang}},
  \bibinfo {author} {\bibfnamefont {Y.-Q.}\ \bibnamefont {Cheng}}, \bibinfo
  {author} {\bibfnamefont {Y.}~\bibnamefont {Yue}}, \bibinfo {author}
  {\bibfnamefont {X.}~\bibnamefont {Han}}, \bibinfo {author} {\bibfnamefont
  {Z.}~\bibnamefont {Zhang}}, \bibinfo {author} {\bibfnamefont
  {Z.}~\bibnamefont {Shan}}, \bibinfo {author} {\bibfnamefont {S.~X.}\
  \bibnamefont {Mao}}, \bibinfo {author} {\bibfnamefont {M.}~\bibnamefont
  {Ye}}, \bibinfo {author} {\bibfnamefont {Y.}~\bibnamefont {Yin}}, \ and\
  \bibinfo {author} {\bibfnamefont {E.}~\bibnamefont {Ma}},\ }\href {\doibase
  10.1038/ncomms1021} {\bibfield  {journal} {\bibinfo  {journal} {Nature
  Communications}\ }\textbf {\bibinfo {volume} {1}},\ \bibinfo {pages} {24}
  (\bibinfo {year} {2010})}\BibitemShut {NoStop}%
\bibitem [{\citenamefont {YL}\ \emph {et~al.}(2008)\citenamefont {YL},
  \citenamefont {DM}, \citenamefont {WM}, \citenamefont {YS},\ and\
  \citenamefont {KL}}]{YL2008}%
  \BibitemOpen
  \bibfield  {author} {\bibinfo {author} {\bibfnamefont {L.}~\bibnamefont
  {YL}}, \bibinfo {author} {\bibfnamefont {W.}~\bibnamefont {DM}}, \bibinfo
  {author} {\bibfnamefont {L.}~\bibnamefont {WM}}, \bibinfo {author}
  {\bibfnamefont {L.}~\bibnamefont {YS}}, \ and\ \bibinfo {author}
  {\bibfnamefont {T.}~\bibnamefont {KL}},\ }\href {\doibase
  10.1016/j.ces.2007.09.028} {\bibfield  {journal} {\bibinfo  {journal}
  {Chemical Engineering Science}\ }\textbf {\bibinfo {volume} {63}},\ \bibinfo
  {pages} {195} (\bibinfo {year} {2008})}\BibitemShut {NoStop}%
\bibitem [{\citenamefont {Sneddon}(1965)}]{SNEDDON196547}%
  \BibitemOpen
  \bibfield  {author} {\bibinfo {author} {\bibfnamefont {I.~N.}\ \bibnamefont
  {Sneddon}},\ }\href {\doibase https://doi.org/10.1016/0020-7225(65)90019-4}
  {\bibfield  {journal} {\bibinfo  {journal} {International Journal of
  Engineering Science}\ }\textbf {\bibinfo {volume} {3}},\ \bibinfo {pages}
  {47} (\bibinfo {year} {1965})}\BibitemShut {NoStop}%
\bibitem [{\citenamefont {Hertz}(1896)}]{Hertz}%
  \BibitemOpen
  \bibfield  {author} {\bibinfo {author} {\bibfnamefont {H.}~\bibnamefont
  {Hertz}},\ }\href {https://doi.org/10.1038/055006f0} {\bibfield  {journal}
  {\bibinfo  {journal} {Miscellaneous papers, MacMillan and Co., New York}\
  }\textbf {\bibinfo {volume} {MacMillan and Co.}} (\bibinfo {year}
  {1896})}\BibitemShut {NoStop}%
\bibitem [{\citenamefont {Pathak}\ \emph {et~al.}(2009)\citenamefont {Pathak},
  \citenamefont {Cambaz}, \citenamefont {Kalidindi}, \citenamefont {Swadener},\
  and\ \citenamefont {Gogotsi}}]{PATHAK2009}%
  \BibitemOpen
  \bibfield  {author} {\bibinfo {author} {\bibfnamefont {S.}~\bibnamefont
  {Pathak}}, \bibinfo {author} {\bibfnamefont {Z.~G.}\ \bibnamefont {Cambaz}},
  \bibinfo {author} {\bibfnamefont {S.~R.}\ \bibnamefont {Kalidindi}}, \bibinfo
  {author} {\bibfnamefont {J.~G.}\ \bibnamefont {Swadener}}, \ and\ \bibinfo
  {author} {\bibfnamefont {Y.}~\bibnamefont {Gogotsi}},\ }\href {\doibase
  https://doi.org/10.1016/j.carbon.2009.03.042} {\bibfield  {journal} {\bibinfo
   {journal} {Carbon}\ }\textbf {\bibinfo {volume} {47}},\ \bibinfo {pages}
  {1969} (\bibinfo {year} {2009})}\BibitemShut {NoStop}%
\bibitem [{\citenamefont {Pathak}\ \emph {et~al.}(2013)\citenamefont {Pathak},
  \citenamefont {Mohan}, \citenamefont {Decolvenaere}, \citenamefont
  {Needleman}, \citenamefont {Bedewy}, \citenamefont {Hart},\ and\
  \citenamefont {Greer}}]{Pathak2013}%
  \BibitemOpen
  \bibfield  {author} {\bibinfo {author} {\bibfnamefont {S.}~\bibnamefont
  {Pathak}}, \bibinfo {author} {\bibfnamefont {N.}~\bibnamefont {Mohan}},
  \bibinfo {author} {\bibfnamefont {E.}~\bibnamefont {Decolvenaere}}, \bibinfo
  {author} {\bibfnamefont {A.}~\bibnamefont {Needleman}}, \bibinfo {author}
  {\bibfnamefont {M.}~\bibnamefont {Bedewy}}, \bibinfo {author} {\bibfnamefont
  {A.~J.}\ \bibnamefont {Hart}}, \ and\ \bibinfo {author} {\bibfnamefont
  {J.~R.}\ \bibnamefont {Greer}},\ }\href {\doibase 10.1021/nn402710j}
  {\bibfield  {journal} {\bibinfo  {journal} {ACS Nano}\ }\textbf {\bibinfo
  {volume} {7}},\ \bibinfo {pages} {8593} (\bibinfo {year} {2013})}\BibitemShut
  {NoStop}%
\bibitem [{\citenamefont {Gibson}\ and\ \citenamefont
  {Ashby}(1997)}]{gibson_ashby_1997}%
  \BibitemOpen
  \bibfield  {author} {\bibinfo {author} {\bibfnamefont {L.~J.}\ \bibnamefont
  {Gibson}}\ and\ \bibinfo {author} {\bibfnamefont {M.~F.}\ \bibnamefont
  {Ashby}},\ }\href {\doibase 10.1017/CBO9781139878326} {\emph {\bibinfo
  {title} {Cellular Solids: Structure and Properties}}},\ \bibinfo {edition}
  {2nd}\ ed.,\ Cambridge Solid State Science Series\ (\bibinfo  {publisher}
  {Cambridge University Press},\ \bibinfo {year} {1997})\BibitemShut {NoStop}%
\bibitem [{\citenamefont {Deschamps}\ \emph {et~al.}(2014)\citenamefont
  {Deschamps}, \citenamefont {Margueritat}, \citenamefont {Martinet},
  \citenamefont {Mermet},\ and\ \citenamefont {Champagnon}}]{Deschamps2014}%
  \BibitemOpen
  \bibfield  {author} {\bibinfo {author} {\bibfnamefont {T.}~\bibnamefont
  {Deschamps}}, \bibinfo {author} {\bibfnamefont {J.}~\bibnamefont
  {Margueritat}}, \bibinfo {author} {\bibfnamefont {C.}~\bibnamefont
  {Martinet}}, \bibinfo {author} {\bibfnamefont {A.}~\bibnamefont {Mermet}}, \
  and\ \bibinfo {author} {\bibfnamefont {B.}~\bibnamefont {Champagnon}},\
  }\href {\doibase 10.1038/srep07193} {\bibfield  {journal} {\bibinfo
  {journal} {Scientific reports}\ }\textbf {\bibinfo {volume} {4}},\ \bibinfo
  {pages} {7193} (\bibinfo {year} {2014})},\ \bibinfo {note}
  {25431218[pmid]}\BibitemShut {NoStop}%
\bibitem [{\citenamefont {Mackenzie}(1950)}]{Mackenzie_1950}%
  \BibitemOpen
  \bibfield  {author} {\bibinfo {author} {\bibfnamefont {J.~K.}\ \bibnamefont
  {Mackenzie}},\ }\href {\doibase 10.1088/0370-1301/63/1/302} {\bibfield
  {journal} {\bibinfo  {journal} {Proceedings of the Physical Society. Section
  B}\ }\textbf {\bibinfo {volume} {63}},\ \bibinfo {pages} {2} (\bibinfo {year}
  {1950})}\BibitemShut {NoStop}%
\bibitem [{\citenamefont {Yu}\ \emph {et~al.}(2005)\citenamefont {Yu},
  \citenamefont {Song}, \citenamefont {Zhang},\ and\ \citenamefont
  {Yang}}]{Yu2005}%
  \BibitemOpen
  \bibfield  {author} {\bibinfo {author} {\bibfnamefont {R.}~\bibnamefont
  {Yu}}, \bibinfo {author} {\bibfnamefont {H.}~\bibnamefont {Song}}, \bibinfo
  {author} {\bibfnamefont {X.-F.}\ \bibnamefont {Zhang}}, \ and\ \bibinfo
  {author} {\bibfnamefont {P.}~\bibnamefont {Yang}},\ }\href {\doibase
  10.1021/jp050973r} {\bibfield  {journal} {\bibinfo  {journal} {The Journal of
  Physical Chemistry B}\ }\textbf {\bibinfo {volume} {109}},\ \bibinfo {pages}
  {6940} (\bibinfo {year} {2005})}\BibitemShut {NoStop}%
\bibitem [{\citenamefont {Harada}\ and\ \citenamefont
  {Asakura}(1996)}]{Harada1996}%
  \BibitemOpen
  \bibfield  {author} {\bibinfo {author} {\bibfnamefont {Y.}~\bibnamefont
  {Harada}}\ and\ \bibinfo {author} {\bibfnamefont {T.}~\bibnamefont
  {Asakura}},\ }\href@noop {} {\bibfield  {journal} {\bibinfo  {journal}
  {Optics communications}\ }\textbf {\bibinfo {volume} {124}},\ \bibinfo
  {pages} {529} (\bibinfo {year} {1996})}\BibitemShut {NoStop}%
\bibitem [{\citenamefont {Trojek}\ \emph {et~al.}(2012)\citenamefont {Trojek},
  \citenamefont {Chv\'{a}tal},\ and\ \citenamefont {Zem\'{a}nek}}]{Trojek2012}%
  \BibitemOpen
  \bibfield  {author} {\bibinfo {author} {\bibfnamefont {J.}~\bibnamefont
  {Trojek}}, \bibinfo {author} {\bibfnamefont {L.}~\bibnamefont {Chv\'{a}tal}},
  \ and\ \bibinfo {author} {\bibfnamefont {P.}~\bibnamefont {Zem\'{a}nek}},\
  }\href@noop {} {\bibfield  {journal} {\bibinfo  {journal} {J. Opt. Soc. Am.
  A}\ }\textbf {\bibinfo {volume} {29}},\ \bibinfo {pages} {1224} (\bibinfo
  {year} {2012})}\BibitemShut {NoStop}%
\bibitem [{\citenamefont {Chaumet}\ and\ \citenamefont
  {Nieto-Vesperinas}(2000)}]{Chaumet2000}%
  \BibitemOpen
  \bibfield  {author} {\bibinfo {author} {\bibfnamefont {P.~C.}\ \bibnamefont
  {Chaumet}}\ and\ \bibinfo {author} {\bibfnamefont {M.}~\bibnamefont
  {Nieto-Vesperinas}},\ }\href@noop {} {\bibfield  {journal} {\bibinfo
  {journal} {Optics letters}\ }\textbf {\bibinfo {volume} {25}},\ \bibinfo
  {pages} {1065} (\bibinfo {year} {2000})}\BibitemShut {NoStop}%
\bibitem [{\citenamefont {Hotta}\ \emph {et~al.}(2009)\citenamefont {Hotta},
  \citenamefont {Hayashi}, \citenamefont {Nishikata},\ and\ \citenamefont
  {Nagata}}]{Hotta2009}%
  \BibitemOpen
  \bibfield  {author} {\bibinfo {author} {\bibfnamefont {M.}~\bibnamefont
  {Hotta}}, \bibinfo {author} {\bibfnamefont {M.}~\bibnamefont {Hayashi}},
  \bibinfo {author} {\bibfnamefont {A.}~\bibnamefont {Nishikata}}, \ and\
  \bibinfo {author} {\bibfnamefont {K.}~\bibnamefont {Nagata}},\ }\href@noop {}
  {\bibfield  {journal} {\bibinfo  {journal} {Isij International - ISIJ INT}\
  }\textbf {\bibinfo {volume} {49}},\ \bibinfo {pages} {1443} (\bibinfo {year}
  {2009})}\BibitemShut {NoStop}%
\end{thebibliography}%
\end{document}